\documentclass{emulateapj}
\pdfoutput=1
\usepackage{amsmath}
\usepackage{amsfonts}
\usepackage{amssymb}
\usepackage{bm}
\usepackage{cases}
\usepackage{float}
\usepackage{graphicx}
\usepackage{latexsym}
\usepackage{longtable}
\usepackage{multirow}
\usepackage{txfonts}
\usepackage{tipa}

\begin{document}

\title{Chemical and Kinematic Properties of the Galactic Disk from the LAMOST and Gaia Sample Stars}

\author{Yepeng Yan\altaffilmark{1}, Cuihua Du\altaffilmark{2}, Shuai Liu\altaffilmark{3}, Hefan Li\ \altaffilmark{1}, Jianrong Shi\altaffilmark{3,2},Yuqin Chen\altaffilmark{3,2}, Jun Ma\altaffilmark{3,2}, Zhenyu Wu\altaffilmark{3,2}}

\affil{$^{1}$School of Physical Sciences, University of Chinese Academy of Sciences, Beijing 100049, China\\ 
$^{2}$College of Astronomy and Space Sciences, University of Chinese Academy of Sciences, Beijing 100049, China; ducuihua@ucas.ac.cn \\ 
$^{3}$Key Laboratory of Optical Astronomy, National Astronomical Observatories, Chinese Academy of Sciences, Beijing 100012, China\\
}

\begin{abstract}

\par We determined the chemical and kinematic properties of the Galactic thin and thick disk using a sample of 307,246 A/F/G/K-type giant stars from the LAMOST spectroscopic survey and Gaia DR2 survey.  Our study found that the thick disk globally exhibits no metallicity radial gradient, but the inner disk ($R \le 8$ kpc) and the outer disk ($R>8$ kpc) have different gradients when they are studied separately. The thin disk also shows two different metallicity radial gradients for the inner disk and the outer disk, and has steep metallicity vertical gradient of d[Fe/H]/d$|z|$ $=-0.12\pm0.0007$ dex kpc$^{-1}$, but it becomes flat when it is measured at increasing radial distance, while the metallicity radial gradient becomes weaker with increasing vertical distance.  Adopting a galaxy potential model, we derived the orbital eccentricity of sample stars and found a downtrend of average eccentricity with increasing metallicity for the thick disk.  The variation of the rotation velocity with the metallicity shows a positive gradient for the thick disk stars and a negative one for the thin disk stars.  Comparisons of our observed results with models of disk formation suggest that radial migration could have influenced the chemical evolution of the thin disk.  The formation of the thick disk could be affected by more than one processes: the accretion model could play an indispensable role, while other formation mechanisms,  such as the radial migration or heating scenario model  could also have a contribution.

\end{abstract}


\section{Introduction}

\par  Our Milky Way Galaxy has been suggested that the disk can be divided into two components: the thin disk and the thick disk, as introduced \cite{Reid83}.
The two components differ in their spatial distribution, metallicity and kinematics.  In the spatial distribution, the range of scale-height and length for the thin disk are about 200-369 pc and 1.00-3.7 kpc, whereas the thick disk has a scale-height of 600-1000 pc and a scale-length of 2.0-5.5 kpc \citep{Du03, Du06, Bilir06, Bilir08, Karaali07,Juric08,Ya10, Chang11, Jia14, Chen17, Wan17}, but these results are still contentious.   
The metallicity gradients of two disks have been studied in previous works based on a variety of samples.
For example, in the radial direction, \cite{Recio-Blanco14} found no metallicity radial gradient for
the thick disk while \cite{Coskunoglu12} and \cite{Li18} derived a positive metallicity gradient. For the thin disk, most of the researchers found a negative  metallicity radial gradient \citep[e.g.,][]{Bilir12, Boeche13}.
In the vertical direction, \cite{Chen11}, \cite{Carrell12}, \cite{Li17}, and \cite{Guctekin17} found a negative metallicity gradient of the thick disk, while \cite{Katz11} and \cite{Li18} derived a flatter gradient.  For the thin disk, a negative metallicity gradient was also found in the vertical direction \citep{Bilir12,Duong18}, while \cite{Mikolaitis14} derived a flatter gradient.
In addition,  some studies also shown there existed an obvious relationship of the rotation velocity with the metallicity for the disk stars. \cite{Lee11} and \cite{Adibekyan13} shown that this relationship is different for the thin and thick disk components. 
In a word, the consentaneous properties of the thick disk population have been characterized by an older population \citep[e.g.,][]{Gilmore88, Bensby14}, enriched in [$\alpha$/Fe], metal-poor \citep{Prochaska00, Lee11, Fuhrmann17}, and higher velocity dispersions \citep{Chiba00} stellar populations compared with the typical thin disk population.  
These different properties imply that two populations may have different formation and evolutionary histories.

\par 
Some models of formation have been proposed for the thick disk, such as the following scenarios: a) accretion \citep{Abadi03}, b) heating  \citep[e.g.,][]{Quinn93, Kazantzidis08, Villalobos08,Villalobos10}, c) radial migration \citep[e.g.,][]{Sellwood02, Binney09, McMillan17}, and d) gas-rich merger \citep[e.g.,][]{Brook04, Brook05}.  These models predict different trends between the kinematics properties and metallicity of disk stars, as well as between their kinematics and spatial distributions.    
For example,  models of disk heating via satellite mergers or a growing thin disk can induce a notable increase in the mean rotation and velocity dispersions of the thick disk stars than the initial thick disk that not affected by these models \citep[][]{Villalobos10},  models of Gas-rich merger predict a rotational velocity gradient
with Galactocentric distance for disks stars near the solar radius
\citep[][]{Brook07}.   \citet{Sales09} shown that the
distribution of orbital eccentricities for nearby thick disk stars
could provide constraints on these proposed formation models. 
\cite{Jing16} determined a preferential scenario by comparing the orbital eccentricities distribution with these simulations, and their results agree with gas-rich merger model of thick disk formation.  So the properties of metallicity and kinematics of the thick disk could provide strong constraints on formation. 

\par 
To study the chemical and kinematic characteristics of the disks, it is necessary to separate the thick disk from thin disk. 
In general, there are mainly four ways to distinguish the local thick and thin disk stars at the solar annulus \citep{Adibekyan11}: a pure kinematical approach \citep{Bensby03}, a pure chemical approach \citep[e.g.,][]{Lee11, Duong18}, a combination of metallicities and spatial distribution \citep{Jing16}, and a combination of kinematics, metallicities, and stellar ages \citep{Haywood08}.   Recently, some researches found a gap in the [$\alpha$/Fe] versus [Fe/H] plane for stars sample \citep[e.g.,][]{Lee11, Bovy16, Duong18, Liu18}, and it is widely used to separate the thick disk stars from the thin disk stars. The thin disk is often defined as the low-[$\alpha$/Fe] population and high-[$\alpha$/Fe] population for the thick disk.   To understand the formation of the Galaxy components, we need more accurate chemical and kinematics information of a large number of stars.  
The large-scale spectroscopic surveys make it possible by providing ideal stellar atmospheric parameters such as metallicity and surface gravity.
The ongoing Large Sky Area Multi-Object
Fiber Spectroscopic Telescope survey \citep[LAMOST, also called Guoshoujing
Telescope;][]{Zhao12} has released more than five
millions stellar spectra with stellar parameters in the DR5 catalog. 
The kinematic studies need to use accurate proper motions and parallaxes with sufficiently small uncertainties. 
The second Gaia data release of Gaia survey \citep{Gaia18a, Gaia18b} provide an unprecedented sample of precisely and accurately measured source. 
These data sets will provide a vast resource to study details of the velocity distribution and give constraints on the formation of the Galactic disk.

\par 
In this work, we have used data from the LAMOST spectroscopic survey and Gaia (DR2) survey to study chemistry and kinematics of the Galactic Disk. The paper is structured as follows. Section 2 introduces the observation data from LAMOST and Gaia, determines the distance and velocity of sample stars, and describes the sample selection. Section 3 presents the result of metallicity and  $\alpha$-abundance variation with radial distance and vertical height. Section 4 investigates the distribution of orbital eccentricities and its trends with metallicity and vertical height and presents the result of rotational velocity variation with radial distance, vertical height, and metallicity. Section 5 discuss the selection bias of our sample, and the formation and evolution of the thin disk and thick disk using our observed results. The summary and conclusions are given in Section 6.

\section{Data}
\subsection{LAMOST and Gaia}
\par 
The LAMOST is also called the GuoShouJing Telescope located at Xinglong Station of the National Astronomical Observatories, Chinese Academy of Sciences (NAOC). It is a reflecting Schmidt telescope with effective aperture of 3.6 m-4.9 m and 4000 fibers within a field of view of $5^\circ$. The LAMOST spectrograph has a resolution of R $\sim$ 1,800 and observed wavelength range spans 3,700{\AA} $\sim$ 9,000 {\AA} \citep{Cui12, Zhao12}. Its observable sky covers $-10^\circ \sim +90^\circ$ declination and the survey reaches a limiting magnitude of $r = 17.8$ (where $r$ denotes magnitude in the SDSS $r$-band). In June 2017, the LAMOST has completed 5 years of survey operation, and in December 2017, the fifth LAMOST data was released, LAMOST DR5. It contains more than 9 million spectra in total. Of these, $\sim 5.34$ million are A/F/G/K-type stars with estimated stellar atmospheric parameters as well as radial velocities, which provides a powerful basic data for astronomers to study the structure and formation of the Milky Way Galaxy.  For example, based on LAMOST DR3 data,  \cite{Liu17}, \cite{Xu18} and \cite{Wang18} derived the stellar density profiles of the Milky Way and spatial structure in the outer disc.  \cite{Li2016, Li2019} used LAMOST M giants to study the Sagittarius stream. 
\par
In this work, the stellar atmospheric parameters ([Fe/H], [$\alpha$/Fe] and surface gravity) are from LSS-GAC DR4 catalog. LSS-GAC (LAMOST Spectroscopic Survey of the Galactic Anticentre) is a major component of the LAMOST Experiment for Galactic Understanding and Exploration.  LSS-GAC Stellar Parameter Pipeline at Peking University [LSP3] \citep{Xiang15, Xiang17a} determines atmospheric parameters by template matching with the MILES spectral library \citep{Sanchez-Blazquez06}. The MILES spectral library consists of 985 stars spanning wide range of stellar atmospheric parameters. The wavelength coverage of the spectra is 3525-7410 {\AA}, and the spectral resolution is about 2.5 {\AA}. The latter is close to that of the LAMOST spectra. The current implementation of LSP3 has achieved an accuracy of 150 K, 0.25 dex, 0.15 dex for the effective temperature, surface gravity, and metallicity, respectively, for the LSS-GAC spectra of F/G/K-type stars of SNRs per pixel higher than 10 \citep{Xiang15}. Values of $\alpha$-element (Mg, Si, Ca and Ti) to iron abundance ratio [$\alpha$/Fe] has also been derived with LSP3, with precisions similar to those achieved by the APOGEE survey for  the giant stars \citep{Xiang17b}. In addition, the radial velocity is from LAMOST DR5 catalog. The determination of the radial velocity makes use of the ELODIE library \citep{Wu11}. 
\par 
The Gaia is a space-based mission mainly on astrometry launched by the European Space Agency (ESA) in December 2013  \citep{Gaia Collaboration16}. 
Gaia mission has already released its second set of data, the Gaia DR2, which provides accurate positions, parallaxes and proper motions for $1.3$ billion sources brighter than magnitude $G \sim 21$ mag, and line-of-sight velocities for 7.2 million stars brighter than GRVS $= 12$ mag \citep{Gaia18a, Gaia18b}. The median uncertainty for the bright sources ($G<14$ mag) is $0.04$ mas, $0.1$ mas at $G = 17$ mag, and $0.7$ mas at $G = 20$ mag, for the parallax and $0.05, 0.2$, and $1.2$ mas\,yr$^{-1}$ for the proper motions, respectively.  More detailed description about the astrometric content of the Gaia DR2 can be found in \cite{Lindegren18}.

\subsection{Distance and velocity determination}
\par 
In this work, we use the Gaia DR2 proper motion and parallax data and restrict parallax uncertainties ($\varpi_{\rm error}/(\varpi-\varpi_{\rm zp})$) smaller than 20$\%$.  The quantity $\varpi$ denotes stellar parallax, and $\varpi_{\rm zp}$ is the global parallax zero-point of Gaia observations. \cite{Butkevich17} confirms that due to various instrumental effects of the Gaia satellite, in particular, to a certain kind of basic-angle variations, these can bias the parallax zero point of an astrometric solution derived from observations. This global parallax zero-point was determined in \cite{Lindegren18} based on observations of quasars: $\varpi_{\rm zp}=-0.029$ mas. Thus, it is necessary to subtract parallax zero-point ($\varpi_{\rm zp}$) when parallax is used to calculate astrophysical quantities \citep{Li19}. We also restrict the error of proper motion in right ascension (pmra$\_$err) and in declination direction (pmdec$\_$err) smaller than $0.2$ mas ${\rm yr^{-1}}$.  We select giant stars by restricting $0 <$ log($g$) $< 3.5$, and restrict these stars with S/N $> 20$ in the $g$-band, radial velocity uncertainties smaller than $10$ ${\rm km\ s^{-1}}$, and error of [Fe/H] smaller than $0.2$ dex.  In total, we obtain 307,246 giant stars.

\par The Gaia DR2 provides precise position and parallax for an unprecedented number of objects, and some astrophysical quantities such as distance and velocities can be inferred using those data.  This is an important task to derive distance and velocities, especially when parallaxes are involved because the effects of the observational errors on the parallaxes and the proper motions  can lead to potentially strong biases \citep{Luri18}. By comparison of the heliocentric distance computed by inverting the Gaia DR2 parallax with derived by \cite{Bailer-Jones18} using a geometrical distance prior for our sample (parallax uncertainties smaller than $20\%$ ), we found that the heliocentric distance of stars are determined precisely just by inverting the parallax ($1/(\varpi-\varpi_{\rm zp})$) for nearby ($1/(\varpi-\varpi_{\rm zp}) < 2$ kpc) sample stars. 
\begin{figure}[]
	\includegraphics[width=1.0\hsize]{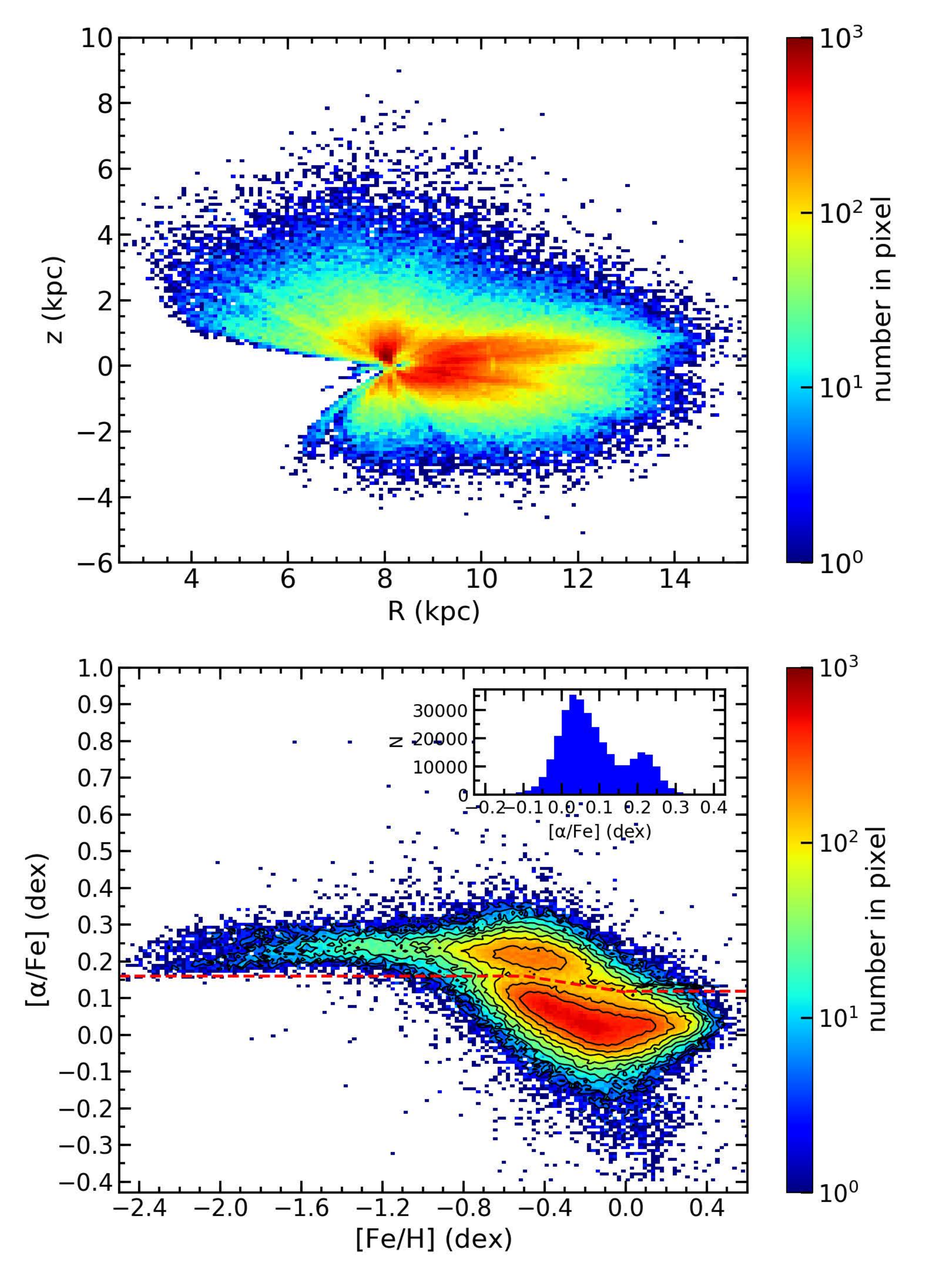}
	\caption{Top panel: space distribution in cylindrical Galactic coordinates for 307,246 giant stars. Bottom panel: [$\alpha$/Fe] as a function of [Fe/H] for these giant stars. Two distinct abundance sequences are found. The red dashed line is dividing line between the high and low-[$\alpha$/Fe] populations. And the stars above red line define the high-[$\alpha$/Fe] stars, while the stars below red line define the low-[$\alpha$/Fe] stars. The inset in the bottom panel shows the histogram of the [$\alpha$/Fe] distribution, where the two populations appear to separate.}
	\label{figure1}
\end{figure}
We discuss on how we determine distances and velocities of stars for nearby ($1/(\varpi-\varpi_{\rm zp}) < 2$ kpc) and distant ($1/(\varpi-\varpi_{\rm zp}) \geq 2$ kpc) stars. 
\subsubsection{The nearby sample stars} 
\begin{figure}[]
	\includegraphics[width=1.0\hsize]{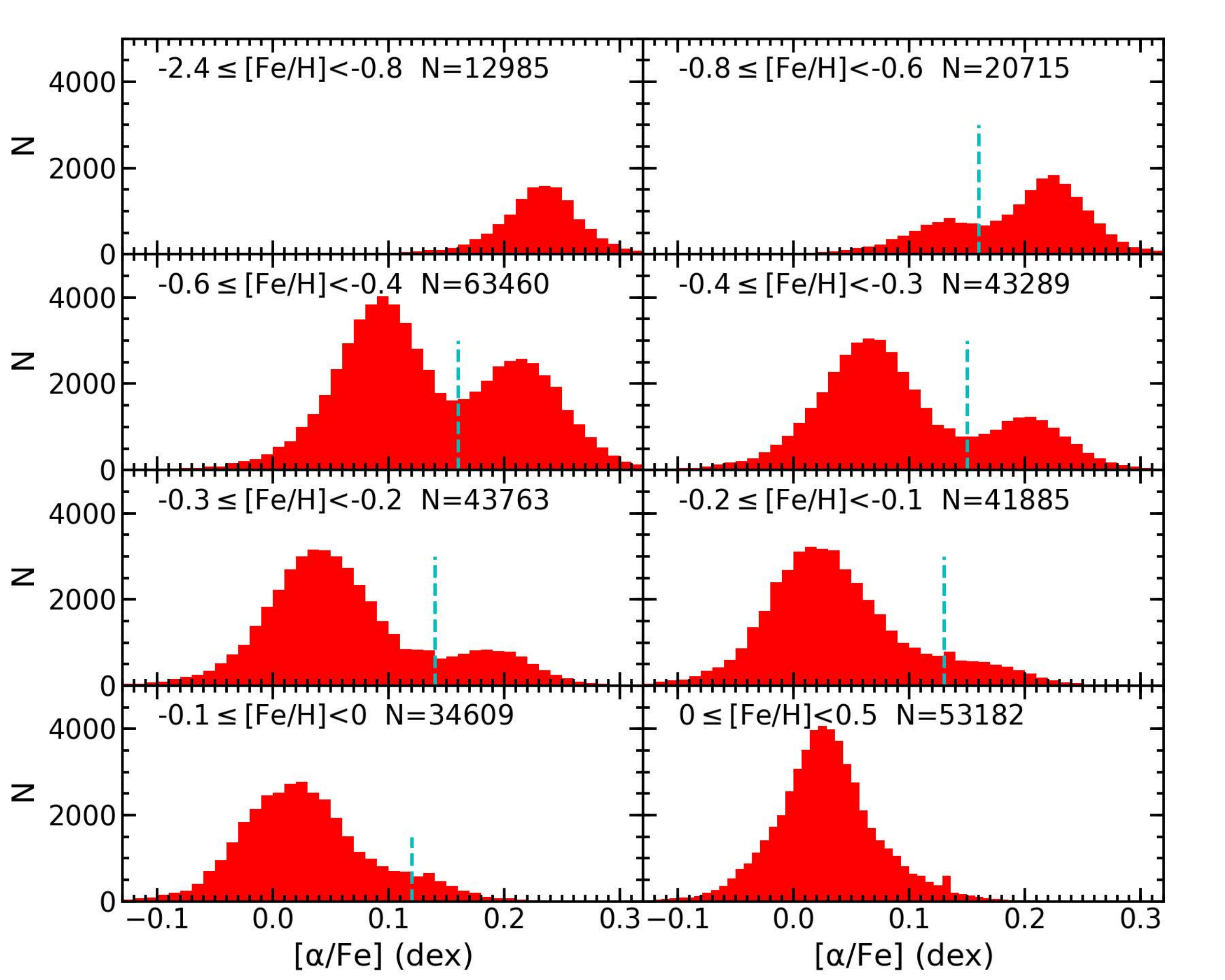}
	\caption{The distribution of [$\alpha$/Fe] in eight metallicity intervals for 307,246 giant stars. The position separating the high-[$\alpha$/Fe] and the low-[$\alpha$/Fe] population is marked with a cyan dash line. }
	\label{figure2}
\end{figure}
\par 
For our sample, 143,303 out of 307,246 ($\sim 46.6\%$) are nearby ($1/(\varpi-\varpi_{\rm zp}) < 2$ kpc) sample stars whose heliocentric distance are computed by inverting the parallax from Gaia DR2: $d=1/(\varpi-\varpi_{\rm zp})$. We then transform the Galactic coordinates $(l, b)$ and heliocentric distance for the stars into a Cartesian Galactocentric coordinate system $(x, y, z)$, and derive the projected distance from the Galactic center using coordinate transformations \citep{Bond10}:
\begin{align}
& x =  R_\odot - d\,\cos(l)\,\cos(b) \nonumber \\
& y =  -d\,\sin(l)\,\cos(b) \\
& z =  d\,\sin(b), \nonumber
\end{align}
Here, we adopt the distance from the Sun to Galactic center is $R_{\odot} = 8.2$ kpc \citep{Bland-Hawthorn16}. $d$ is the distance from the star to the Sun, and $l$ and $b$ are the Galactic longitude and latitude. The proper motions together with the radial velocity are used to derive Galactic velocity components $(U, V, W)$, and Galactocentric cylindrical components $V_{\varphi}$ and their error. Here we adopt a Local Standard of Rest velocity $V_{{\rm LSR}} = 220\ {\rm km s^{-1}}$, and the solar peculiar motion $(V_x^{\odot,{\rm pec}},V_y^{\odot,{\rm pec}},V_z^{\odot,\rm pec}) = {\rm (10.0\ km s^{-1}, 11.0\ km s^{-1}, 7.0\ km s^{-1})}$ \citep{Tian2015, Bland-Hawthorn16}. 

\subsubsection{The distant sample stars}
\begin{figure}[]
	\includegraphics[width=1.0\hsize]{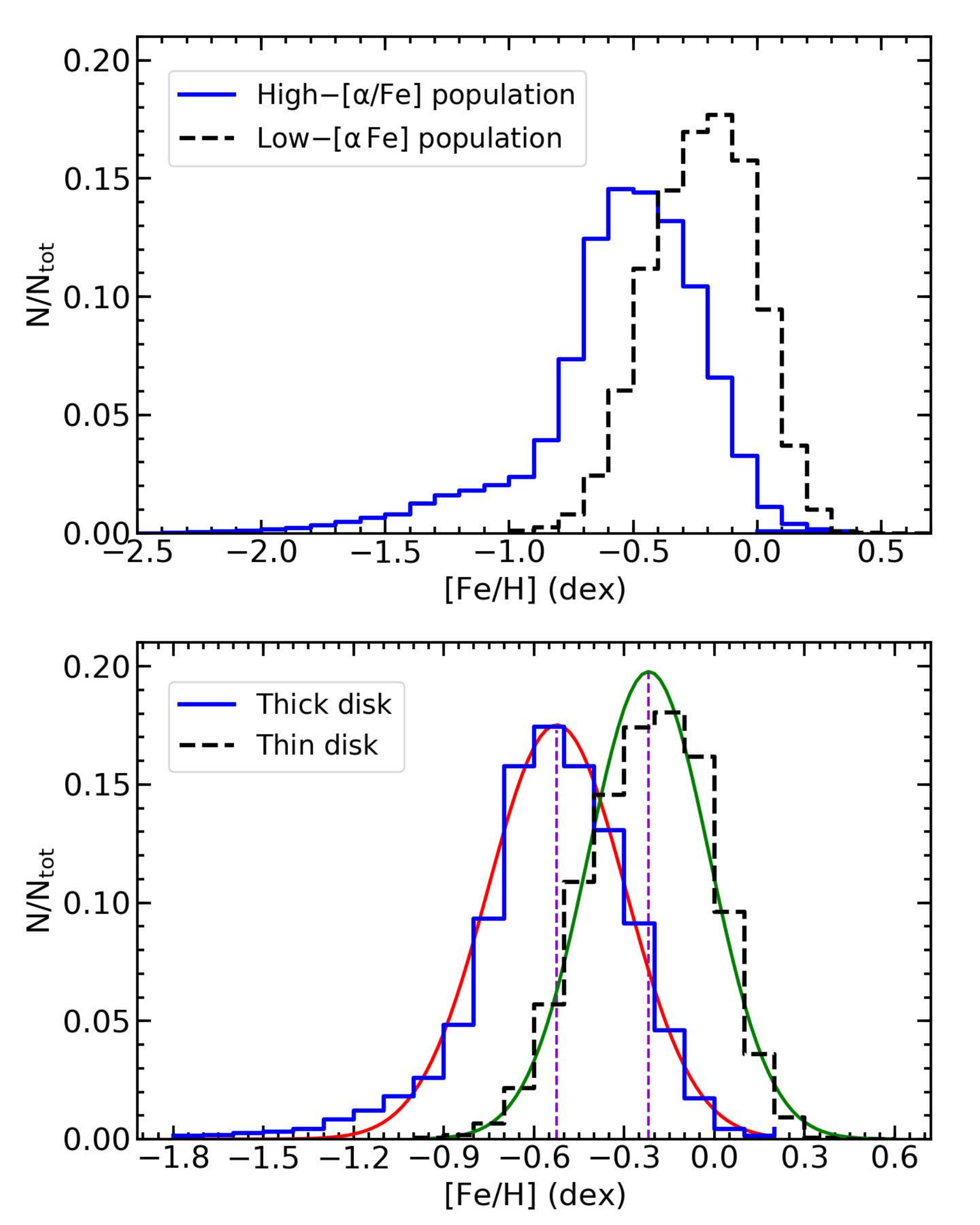}
	\caption{Top panel: [Fe/H] distribution for the high${-}{\rm[\alpha/Fe]}$ and low${-}{\rm[\alpha/Fe]}$ populations. Bottom panel: [Fe/H] distribution for the thin disk and thick disk components selected according to a gap in the [$\alpha$/Fe] versus [Fe/H] plane and kinematics. They can be described by two Gaussian model with peaks at [Fe/H] $\sim -0.21$ and standard deviation $\sigma_{\rm [Fe/H]} \sim 0.20$ for the thin disk, and [Fe/H] $\sim -0.52$, $\sigma_{\rm [Fe/H]} \sim 0.23$ for the thick disk stars with [Fe/H] $>$ -1.2 dex. But the metallicity  functions of the global thick disk are not Gaussian, and it has an extended metallicity tail.}
	\label{figure3}
\end{figure}
\par 
For 163,943 out of 307,246 ($\sim53.4\%$) distant ($1/(\varpi-\varpi_{\rm zp}) \geq 2$ kpc) sample stars, we use the Bayesian approach following \cite{Bailer-Jones15}, \cite{Astraatmadja16a, Astraatmadja16b} and \cite{Luri18} to determine the stellar distance and velocity. According to Bayes formula, the posterior probability $ P(\bm{{\rm \theta}}|\bm{{\rm x}})$ of observed star can be obtained. \bm{${\rm x}$} and  \bm{${\rm \theta}$} are observed data vector and parameters vector we expect to obtain, respectively. The posterior probability $ P(\bm{{\rm \theta}}|\bm{{\rm x}})$ denotes the probability distribution of the parameters under given observed data. The data vector are the parallax ($\varpi$), proper motion in right ascension ($\mu_{\alpha^*}$) and declination ($\mu_{\delta}$), written as the column vector
\begin{eqnarray}
\bm{{\rm x}} = (\varpi-\varpi_{\rm zp}, \mu_{\alpha^*}, \mu_{\delta})^{\rm T}
\end{eqnarray}
The symbol `T' stands for matrix transpose. The data vector (\bm{${\rm x}$}) has units mas, mas ${\rm yr^{-1}}$ and mas ${\rm yr^{-1}}$,  respectively, and has a covariance matrix as follows:
\begin{align}
{\rm \Sigma}=
\begin{pmatrix} \sigma^2_{\varpi} & \sigma_{\varpi}\sigma_{ \mu_{\alpha^*}}\rho(\varpi,\mu_{\alpha^*}) & \sigma_{\varpi}\sigma_{\mu_{\delta}}\rho(\varpi,\mu_{\delta}) \\ \sigma_{\varpi}\sigma_{ \mu_{\alpha^*}}\rho(\varpi,\mu_{\alpha^*}) & \sigma^2_{\mu_{\alpha^*}} & \sigma_{ \mu_{\alpha^*}}\sigma_{\mu_{\delta}}\rho(\mu_{\alpha^*},\mu_{\delta}) \\ \sigma_{\varpi}\sigma_{\mu_{\delta}}\rho(\varpi,\mu_{\delta}) & \sigma_{ \mu_{\alpha^*}}\sigma_{\mu_{\delta}}\rho(\mu_{\alpha^*},\mu_{\delta}) & \sigma_{\mu_{\delta}}^2  \end{pmatrix} 
\end{align}
$\rho(i,j)$ denotes the correlation coefficient between the astrometric parameters $i$ and $j$. $\sigma_k$ is the error of astrometric parameters $k$. The parameters vector are the heliocentric distance ($d$), tangential speed ($v$), and travel direction ($ \phi$, increasing anti-clockwise from North), written as
\begin{eqnarray}
\bm{{\rm \theta}} = (d, v, \phi)^{\rm T}
\end{eqnarray}
with units pc, km ${\rm s^{-1}}$ and radians, respectively.
If there is no error in data vector, components of the parameters vector \bm{${\rm \theta}$}  are given by the simple geometrical transformation
\begin{eqnarray}
\begin{cases}
d=\frac{10^3}{(\varpi-\varpi_{\rm zp})} \\v=4.74 \frac{\sqrt{\mu_{\alpha^*}^2+\mu_{\delta}^2}}{\rm mas\cdot yr^{-1}} \frac{1}{(\varpi-\varpi_{\rm zp})\cdot {\rm kpc}}{\rm km\ s^{-1}}\\ 
\phi=\arctan(\frac{\mu_{\alpha^*}}{\mu_{\delta}})
\end{cases}
\end{eqnarray}
\par 
The 3D posterior over the parameters \bm{${\rm \theta}}$ is according to the Bayes formula
\begin{eqnarray}
P(\bm{{\rm \theta}}|\bm{{\rm x}})\propto P(\bm{{\rm x}}|\bm{{\rm \theta}})P(\bm{{\rm \theta}})
\end{eqnarray}
where $P(\bm{{\rm x}}|\bm{{\rm \theta}})$ is called the likelihood probability. The likelihood probability usually represents a adopted model. $\bm{{\rm \theta}}$ represents parameters of this model. The likelihood probability $P(\bm{{\rm x}}|\bm{{\rm \theta}})$ is a multidimensional Gaussian distribution centered on $\bm{{\rm m}}$, 
\begin{eqnarray}
\bm{{\rm m}}= (\frac{10^3}{d}, {c_2} \frac{10^3v  \sin\phi}{d}, {c_2} \frac{10^3v  \cos\phi}{d})^{\rm T}
\end{eqnarray}
where $c_2 = ({\rm pc\cdot mas\cdot yr^{-1} })/({\rm 4.74\cdot km\ s^{-1}})$, it is a result of unit of astrophysical quantities conversion.  And, $\bm{{\rm m}}$ represents a set of theoretical values predicted by this  model. The likelihood probability  can be written as,
\begin{eqnarray}
P(\bm{{\rm x}}|\bm{{\rm \theta}})\propto \textrm{exp}[-\frac{1}{2}(\bm{{\rm x}}-\bm{{\rm m}}(\bm{{\rm \theta}}))^{\rm T}\Sigma^{-1}(\bm{{\rm x}}-\bm{{\rm m}}(\bm{{\rm \theta}}))]
\end{eqnarray}
Here a separable prior distribution is used \citep{Luri18}
\begin{eqnarray}
P(\bm{{\rm \theta}})=P(d)P(v)P(\phi)
\end{eqnarray}
with
\begin{align}
P(d) &\propto \begin{cases} d^2e^{-d/L(a,b)}& d>0\\ 0& d\le 0 \end{cases}\\
P(v) &\propto \begin{cases} (\frac{v}{v_{\rm max}})^{\alpha-1}(1-\frac{v}{v_{\rm max}})^{\beta-1}& {\rm if}\ 0 \le v \le v_{\rm max}\\ 0& \text{otherwise} \end{cases}\\
P(\phi) &\propto \frac{1}{2\pi}
\end{align}
The prior distribution of distance is the exponentially decreasing space density introduced in \cite{Bailer-Jones15}. And we adopt the length scale of Galactic longitude and latitude dependent \citep{Bailer-Jones18}, $L(a,b)$, which is obtained by fitting a spherical harmonic model. The prior over speed is a beta distribution, and we adopt $\alpha=2$, $\beta=3$ and $ v_{\rm max}=750\ {\rm km\ s^{-1}}$. The prior over the angle $\phi$ is uniform. 
\begin{figure}[]
	\includegraphics[width=1.0\hsize]{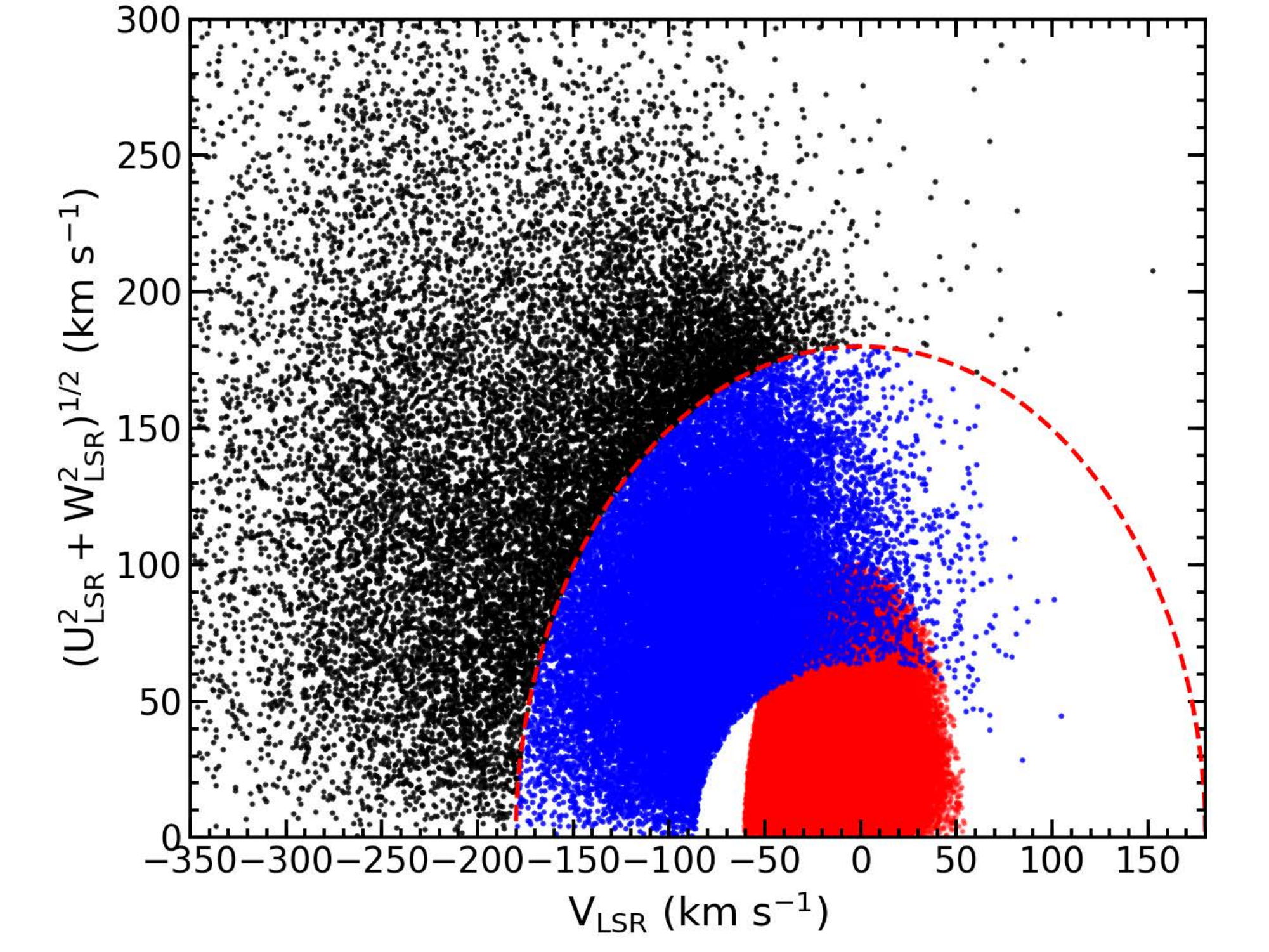}
	\caption{Toomre diagram for our sample stars. the red dashed line shows the values of the total spatial velocity  $v_{{\rm tot}} = \sqrt{U_{{\rm LSR} } + V_{{\rm LSR}} + W_{{\rm LSR}} }= 180$ ${\rm km\ s^{-1}}$. The thick disk stars are represented by blue dots and the thin disk stars are red dots. Those stars with $v_{{\rm tot}} > 180$ ${\rm km\ s^{-1}}$ may be halo stars, which are marked by black dots.}
	\label{figure4}
\end{figure}
\par 
Although the 3D posterior probability distribution $ P(\bm{{\rm \theta}}|\bm{{\rm x}})$ can be obtained using Eq. (6), Eq. (8) and Eq. (9), the posterior probability does not has a simple form. Thus we characterize it by using Markov chain Monte Carlo (MCMC) sampler EMCEE \citep{Goodman10, Foreman-Mackey13}. We run each chain using 100 walkers and 100 steps, for a total of 10000 random samples drawn from the posterior distribution ($ P(\bm{{\rm \theta}}|\bm{{\rm x}})$). We also sample 10000 random samples for radial velocity and assume uniform priors on radial velocity. The radial velocity and its error are provided by LAMOST catalogue and  don't depend on parallax and proper motion. For a star, we obtain 10000 posterior samples including its heliocentric distance ($d$), tangential speed ($v$), direction of travel ($\phi$) and radial velocity ($rv$). We then directly use these random samples to derive Cartesian Galactocentric coordinate ($x,y,z$), the projected distance from the Galactic center, Galactic velocity components $(U, V, W)$, and Galactocentric cylindrical component $V_{\varphi}$ as described section 2.2.1. Here we assume the same parameters for $R_{\odot}$, $V_{{\rm LSR}}$ and  solar peculiar motion presented in section 2.2.1. We choose median as an estimator of these astrophysical quantities, and standard deviation of the quantities define the uncertainty in the estimate value.

\subsection{Sample selection}
\par The top panel of Figure \ref{figure1} shows that the range of Galactocentric radius covered by our sample of 307,246 giant stars within $4 < R < 15$ kpc, extending up to 6 kpc in height from the Galactic plane. In this work, we combine the kinematics with the chemical abundances ([$\alpha$/Fe] and [Fe/H]) to distinguish local thick disk from the thin disk stars. 
The thick disk stars are believed to have higher [$\alpha$/Fe] ratios and lower [Fe/H] than the thin-disk stars based on several previous element abundance analyses \citep{Bensby03, Bensby05}. Therefore, it is widely used that adopting dividing line in [$\alpha$/Fe] versus [Fe/H] to define high-[$\alpha$/Fe] and low-[$\alpha$/Fe] populations. However, the adopted dividing line between the high and low-[$\alpha$/Fe] populations is different in some studies. The [$\alpha$/Fe]-[Fe/H] distribution is shown in the bottom panel of Figure \ref{figure1}. As shown in this panel, there is a gap that divides sample stars into low-[$\alpha/$Fe] and high-[$\alpha$/Fe] populations. The determination of our dividing line between the high and low-[$\alpha$/Fe] populations follow \cite{Adibekyan11} and \cite{Recio-Blanco14}. As shown in Figure \ref{figure2}, we divide the sample into eight metallicity bins from [Fe/H]= -2.4 to 0.5 and identify the minima in the [$\alpha$/Fe] histograms for each bin. We determine eight separation point in [$\alpha$/Fe] versus [Fe/H] plane according to minima in the [$\alpha$/Fe] histograms for each bin: (-2.4, 0.16), (-0.7, 0.16), (-0.5, 0.16), (-0.35, 0.15), (-0.25, 0.14), (-0.15, 0.13), (-0.05, 0.12), (0.5, 0.12).  The separation curve in the bottom panel of Figure \ref{figure1} is the simple connection of these separation points. High-[$\alpha$/Fe] population  is defined as stars above the separation curve (red line in the bottom panel of Figure \ref{figure1}), while  low-[$\alpha$/Fe] population is stars below the separation curve.  High-[$\alpha$/Fe] population extend from [Fe/H] $\approx$ -2.4 to 0.1 and [Fe/H] $\approx$ -0.9 to 0.4 for low-[$\alpha$/Fe].
\par 
The kinematic approach of defining the thick and thin disk stars proposed by \cite{Bensby03} was adopted by many studies \citep[e.g.,][]{Reddy06, Coskunoglu12, Bensby14, Li17}. This method assumes that the Galactic velocities ($U_{\textrm{LSR}}$, $V_{\textrm{LSR}}$, $W_{\textrm{LSR}}$) have  Gaussian distribution given by the equation \citep{Bensby03}. 
\begin{align}
f(U,V,W)=k\cdot \textrm{exp}(\frac{U_{\textrm{LSR}}^2}{2\sigma_{U}^2}-\frac{(V_{\textrm{LSR}}-V_{\textrm{asym}})^2}{2\sigma_{V}^2}-\frac{W_{\textrm{LSR}}^2}{2\sigma_{W}^2})
\end{align}
where 
\begin{eqnarray}
k=\frac{1}{(2\pi)^{3/2}\sigma_{U}\sigma_{V}\sigma_{W}}
\end{eqnarray}
Here, $\sigma_{U}$, $\sigma_{V}$, and $\sigma_{W}$ are the characteristic velocity dispersions, and $V_{\textrm{asym}}$ is the asymmetric drift, and their values are listed in Table \ref{Table 1} \citep{Bensby03}. $U_{\textrm{LSR}}$, $V_{\textrm{LSR}}$, $W_{\textrm{LSR}}$ are stellar velocity relative to Local Standard of Rest.
\begin{table}[hbp]
\centering
\caption{ \upshape {$X$ is the observed fraction of stars for the populations in the solar
		 neighborhood and $V_{\textrm{asym}}$ is the asymmetric drift, as well as characteristic velocity dispersions ($\sigma_{U}$, $\sigma_{V}$, and $\sigma_{W}$)  are listed. } }
\label{Table 1}
\begin{tabular}{llllll}
\hline
\hline 
&$X$&$\sigma_{U}$& $\sigma_{V}$&$\sigma_{W}$ &$V_{\textrm{asym}}$\\ &  &  &  & [km/s] \\
\hline
Thin disk (D) & 0.94 & 35 & 20& 16 &-15\\
Thick disk (TD)& 0.06& 67& 38& 35 &-46\\
Halo (H) & 0.0015& 160& 90& 90& -220\\
\hline	
\end{tabular}
\end{table}
By dividing the thick disk probability (TD) with the thin disk (D), we obtain the relative probabilities for thick-disk-to-thin-disk (TD/D) as follows:
\begin{eqnarray}
\textrm{TD/D} = \frac{X_{\textrm{TD}}\cdot f_{\textrm{TD}}}{X_{\textrm{D}}\cdot f_{\textrm{D}}}
\end{eqnarray} 
Here, $X$ is the observed fraction of stars for the populations in the solar neighborhood, $X_{\textrm{TD}}$ and $X_{\textrm{D}}$ represent the fraction for the thick disk and the thin disk, respectively. Their values are listed in Table \ref{Table 1}. $f_{\textrm{TD}}$ and $f_{\textrm{D}}$ represent Gaussian distribution of Galactic velocities for the thick and thin disk, and they can be calculated with Eq. (13) and Eq. (14) for a given star with Galactic velocities ($U_{\textrm{LSR}}$, $V_{\textrm{LSR}}$, $W_{\textrm{LSR}}$). TD and D are the probabilities that given stars belong to the thick disk and the thin disk, respectively.
We selected TD/D $> 5$ (implying those stars are five times more likely to be thick disk stars than thin disk stars) from high-[$\alpha$/Fe] population as the thick disk stars, and those TD/D $< 0.2$ from low-[$\alpha$/Fe] population as the thin disk stars. Toomre diagram is also an effective way of displaying the sample, which has been widely used to define halo stars \citep[e.g.,][]{Venn04, Nissen10, Bonaca17}. In this work, we use Toomre diagram to exclude halo stars. Figure \ref{figure4} presents the Toomre diagram of the sample stars, and the dashed line shows the values of the total spatial velocity  $v_{{\rm tot}} = \sqrt{U_{{\rm LSR} } + V_{{\rm LSR}} + W_{{\rm LSR}} }= 180$ ${\rm km\ s^{-1}}$. The thick disk stars are represented by blue dots, thin disk stars are red dots while halo stars are black dots. Also, we limited the disk sample stars within $v_{{\rm tot}} < 180$ ${\rm km\ s^{-1}}$ \citep{Nissen10, Bensby14, Bonaca17}.
\par 
We find that our sample of the thick disk contains few metal-poor stars (645 thick disk stars with -1.8 $<$[Fe/H] $<$ -1.2 dex and 75 stars with -2.4 $<$ [Fe/H] $<$ -1.8 dex), and they account for about 2.4$\%$ of the thick disk sample. It is well known that the halo stars are more metal-poor \citep[e.g.,][]{Carollo07, Carollo10, Du18, Li19} and the existence of metal-weak thick disk (MWTD) has  been confirmed by many works \citep[e.g.,][]{Morrison90,Beers95,Chiba98,Martin98,Chiba00,Beers02,Beers14,Tian19}. So these metal-poor stars may be contaminated by halo stars or, at least partially, are metal-weak thick disk stars. But, we find that these metal-poor stars of the thick disk ([Fe/H] $<-1.2$ dex) have more high mean rotational velocity of $<V_{\phi}>\ \sim 114$ $\rm {km\ s^{-1}}$ than halo stars of the slight spin \citep{Carollo10,Deason17}, and this value is consistent with the result of \cite{Carollo10} in which  mean rotational velocity of the MWTD ([Fe/H]$<$ -0.8 dex) is $125 \pm 4$ $\rm {km\ s^{-1}}$. Therefore， the assumption that our metal-poor sample stars may be contaminated by halo stars can be excluded. Alternatively, they probably belong to the MWTD. \cite{Carollo10} reported that the metallicity range for stars that are likely members of the MWTD is -1.8 $\lesssim$ [Fe/H] $\lesssim$ -0.8 dex, so we exclude 75 stars with [Fe/H]　$<$　-1.8 dex for the thick disk. In summary, our sample of the thick disk contains both the canonical thick disk and the MWTD.
\par 
Here we briefly summarize our selection  criteria of the thin disk and thick disk. We obtained 29,966 thick disk stars with high-[$\alpha$/Fe], [Fe/H] $>-1.8$ dex, TD/D $> 5$, $v_{{\rm tot}} < 180$ ${\rm km\ s^{-1}}$, and 179,092 thin disk stars with low-[$\alpha$/Fe] and TD/D $< 0.2$. The bottom panel of Figure \ref{figure3} indicates the metallicity distribution of the thick and thin disk populations can be described by two Gaussian model with peaks at [Fe/H] $\sim -0.21$ and standard deviation $\sigma_{\rm [Fe/H]} \sim 0.20$ for the thin disk, and [Fe/H] $\sim -0.52$, $\sigma_{\rm [Fe/H]} \sim 0.23$ for the thick disk with [Fe/H] $>$ -1.2 dex. The metallicity  function of the global thick disk has an extended metallicity tail, which indicated the existence of the metal-weak thick disk.

\section{Metallicity and  $\alpha$-Abundance Gradient of the Galactic Disk}
In this section, we use our sample stars to examine the observed gradients of metallicity with $R$ and $|z|$, as well as the gradient of $\alpha$-abundance with $|z|$ for the thin disk and the thick disk populations. Firstly, we introduce the estimation of gradient and its error.

\subsection{Gradient and its error estimation}
\par 
As an example, we only introduce the estimation of metallicity  radial gradient and its error. For a given sample,  the total number of stars is $N$ and each star is arranged in order, numbered $i$ ($i= 1, 2, 3, ..., N$), we consider ${\rm [Fe/H]}_{i}$ as a function of radial distance, $R_i$, and ${\rm [Fe/H]}_{i}$ has a noise $\sigma_i$. The maximum likelihood estimation is used to estimate gradient, and the likelihood function is as follows:
\begin{eqnarray}
P({\rm [Fe/H]}|R,\sigma,k,b) = \prod_{i=1}^{N} \frac{1}{\sqrt{2\pi \sigma^2_{i}}} e^{- \frac{({\rm [Fe/H]}_{i} - kR_{i} -b)^2}{2\sigma^2_{i}}}
\end{eqnarray}
where $k$, $b$, $\sigma$ denote fitting slope, intercept and [Fe/H] error, respectively. The likelihood function is a Gaussian function. The slope and intercept can be evaluated by maximizing the likelihood function. More detailed discussion about fitting slope  can be found in \cite{Hogg10}.
\par 
We adopt the Bayesian approach to estimate the error of the metallicity radial gradient. The data vector includes metallicity, [Fe/H] error, and radial distance. The parameters vector are $k$ and $b$. The posterior over the parameters is
\begin{align}
P(k,b|R,{\rm [Fe/H]}, \sigma) \propto P({\rm [Fe/H]}|R, \sigma, k, b)P(k,b),
\end{align}
where the marginalization is used as a likelihood function, and $P({\rm [Fe/H]}|R, \sigma, k, b, f)$ is given in equation (16).  We use uniform priors 
\begin{align}
P(k,b)=P(k)P(b)
\end{align}
with
\begin{align}
P(k) &\propto \begin{cases} 1 & -1< k <-1\\ 0& {\rm otherwise} \end{cases}\\
P(b) &\propto \begin{cases} 1 & -5< b <5\\ 0& \text{otherwise} \end{cases}
\end{align}
The posterior distribution can be obtained, and we characterize it by using Markov chain Monte Carlo (MCMC) sampler EMCEE. We run 300 burn-in steps to let the walkers explore the parameter space and then run each chain using 100 walkers and 100 steps to get a total of 10000 random gradients from the posterior distribution.  The standard deviation of 10000 random gradients is used to define the uncertainty of gradient in the estimation.

\subsection{Radial gradients of the Galactic disk}
\begin{figure}[]
	\includegraphics[width=1.0\hsize]{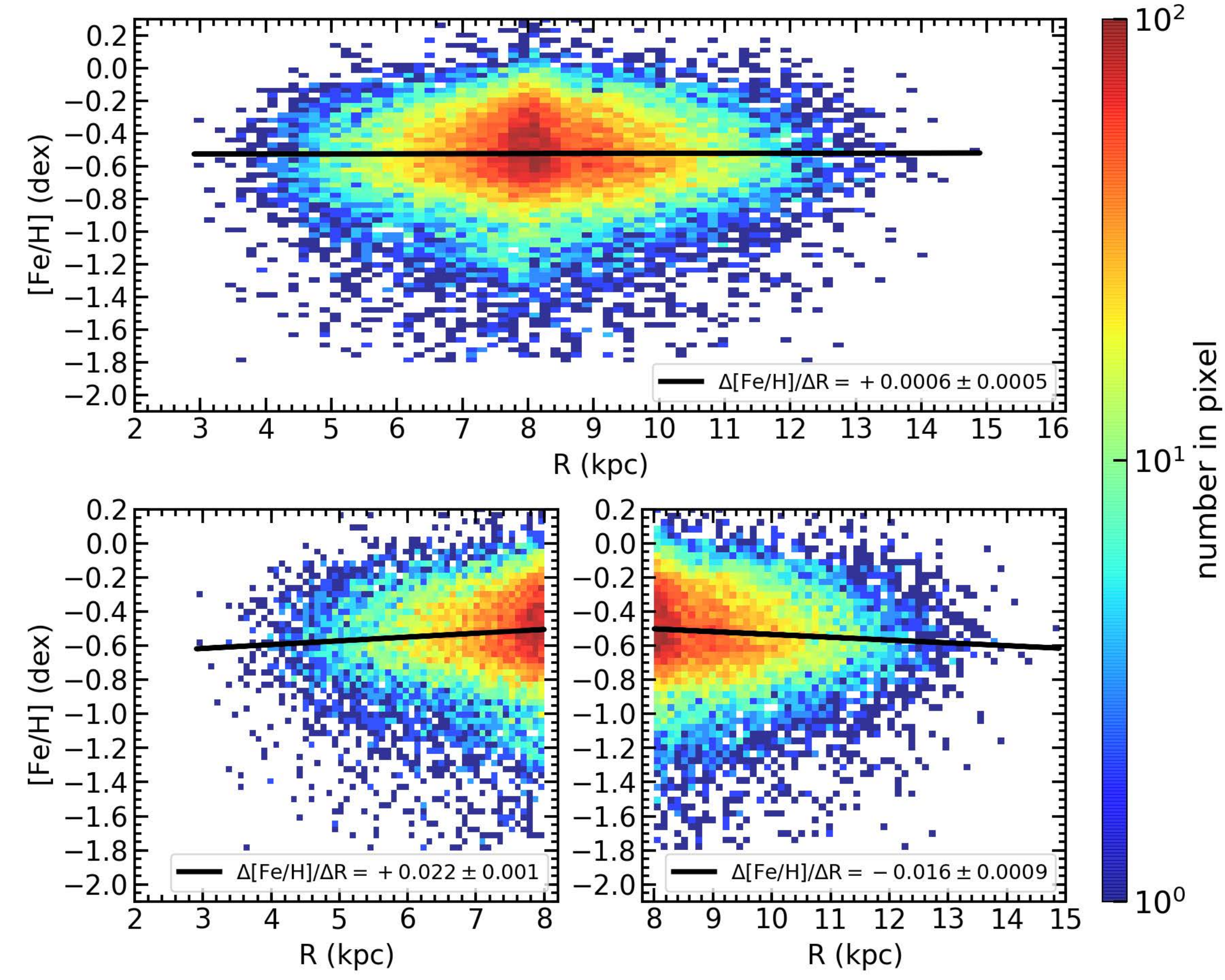}
	\caption{Top panel: metallicity  radial gradient for the thick disk stars.  Bottom panel: metallicity radial gradient for the inner disk ($R \le 8$ kpc) and outer disk ($R > 8$ kpc) of the thick disk. }
	\label{figure5}
\end{figure}
\par 
The metallicity gradient with radial distance for the thick disk stars is given in the top panel of Figure \ref{figure5}. The metallicity radial gradient of the thick disk is ${\rm d[Fe/H]/d}R = +0.0006\pm0.0005$ dex ${\rm kpc^{-1}}$. So our result indicates that the thick disk has a basically flat radial distribution, which is consistent with the result of \cite{Recio-Blanco14}, \cite{Mikolaitis14} and \cite{Peng18} as shown in Table \ref{Table 2}.  However, it is different from what was obtained by \cite{Coskunoglu12}, who used about 17 000 F-type and G-type dwarfs from RAdial Velocity Experiment (RAVE) Data Release 3 (DR3). Also, \cite{Li17, Li18} used 2035 thick-disk giant stars from LAMOST DR3 and disk stars from Apache Point Observatory Galactic Evolution Experiment data release 13 (DR13 hereafter) combined Tycho-Gaia data to derived a flat gradient for the thick disk.
The bottom panel of Figure \ref{figure5} shows several interesting features. 
When we study the variation of metallicity radial gradient with radial distance for the thick disk, we found two different metallicity gradients in the inner ($R \le 8$ kpc) and the outer disk ($R > 8$ kpc). The inner ($R \le 8$ kpc) disk of the thick disk has a positive gradient ${\rm d[Fe/H]/d}R = +0.022\pm0.001$ dex ${\rm kpc^{-1}}$ while the outer disk ($R > 8$ kpc) has a negative gradient ${\rm d[Fe/H]/d}R = -0.016\pm0.0009$ dex ${\rm kpc^{-1}}$. 
\begin{figure}[]
	\includegraphics[width=1.0\hsize]{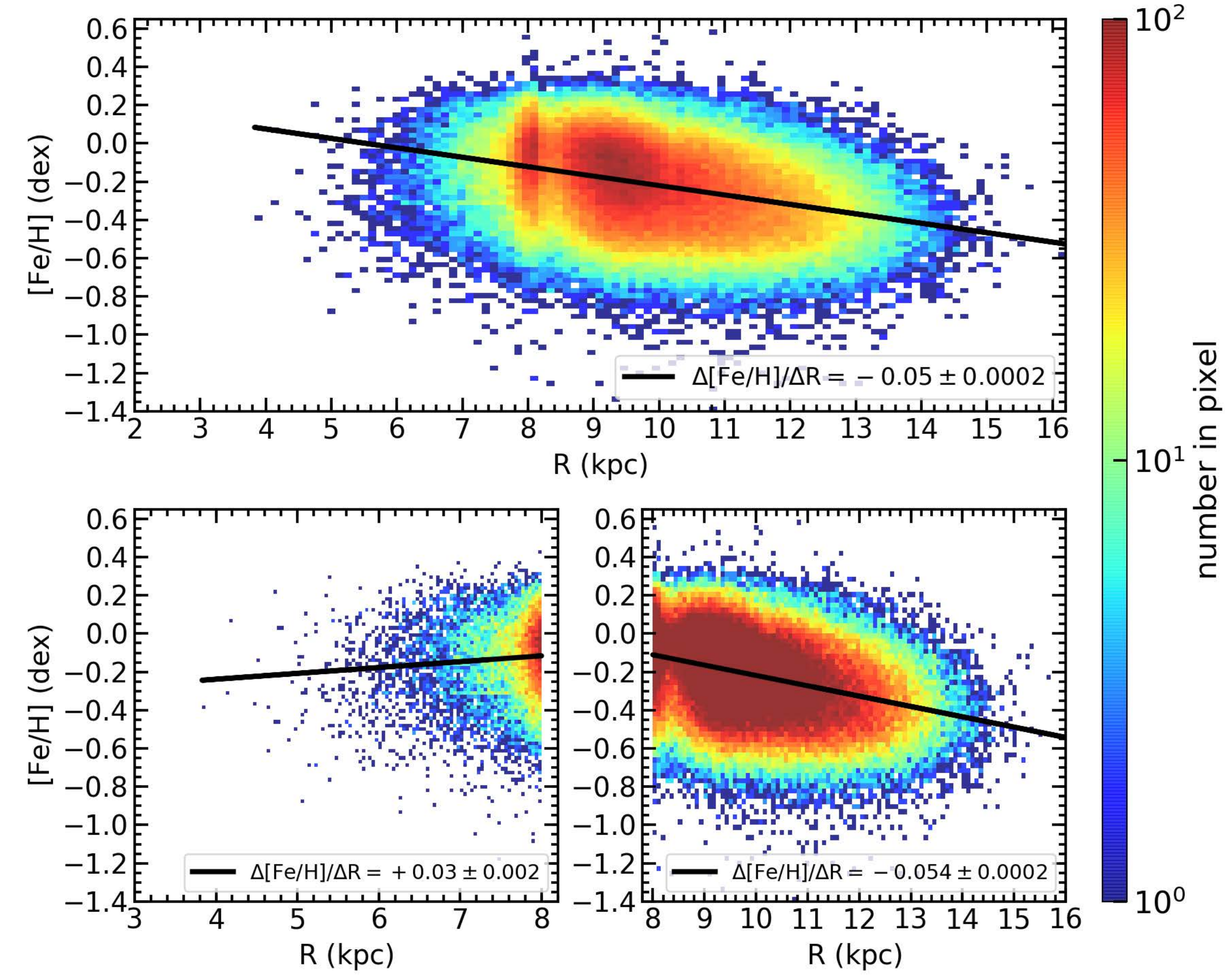}
	\caption{Top panel: metallicity radial gradient for the thin disk stars. Bottom panel: metallicity radial gradient for the inner disk ($R < 8$ kpc) and outer disk ($R > 8$ kpc) of the thin disk. }
	\label{figure6}
\end{figure}
\par 
The top panel of Figure \ref{figure6} shows the metallicity gradient with radial distance for the thin disk stars. Our result shows that the thin disk has a negative  metallicity gradient along the radial direction, ${\rm d[Fe/H]/d}R = -0.05\pm0.0002$ dex ${\rm kpc^{-1}}$. This result is consistent with the results summarized in Table \ref{Table 2} in which the thin disk has a negative  metallicity gradient ($-0.058 \le {\rm d[Fe/H]/d}R \le -0.027$ dex ${\rm kpc^{-1}}$). So the metallicity radial gradient of the thin disk is in good agreement with the results from previous works. It is interesting that two different radial metallicity gradients are also found for the inner ($R \le 8$ kpc) and the outer disk ($R > 8$ kpc) of the thin disk. The inner ($R \le 8$ kpc) disk of the thin disk has a positive gradient ${\rm d[Fe/H]/d}R = +0.03\pm0.002$ dex ${\rm kpc^{-1}}$ while the outer disk ($R > 8$ kpc) has a negative gradient ${\rm d[Fe/H]/d}R = -0.054\pm0.0002$ dex ${\rm kpc^{-1}}$.

\par
\cite{Curir12} reported that,  a positive radial metallicity slope for the inner early Galactic disk ($R \lesssim 10$ kpc), combined with the usual decreasing slope in the outer disk ($R \gtrsim 10$ kpc), in their N-body simulations (including radial migration), plays a critical role to produce a positive rotation-metallicity correlation in the thick disk.  As shown in Table \ref{Table 2}, a positive  rotational velocity gradient with metallicity for the thick disk has been confirmed by many studies, which provides an important evidence for  this hypothetical relationship of the metallicity with radial distance for the early Galactic disk. Furthermore, \cite{Curir14} shown that initial metallicity imprint could not be washed out by secular dynamical processes. This variation trend of metallicity distribution with radial distance for the early Galactic disk is consistent with  our results of the thick and thin disk radial metallicity, only the turning point of radial metallicity slope is different. Our results give the variation trend that the metallicity radial gradient is positive in the inner disk and negative in the outer disk for the thick disk and thin disk, which may be  this  initial metallicity imprint. 

\begin{table*}
	\begin{center}
		\centering
		\caption{ \upshape {The metallicity, [$\alpha$/Fe] and rotational velocity gradients of the thick disk and thin disk in the literatures.} }
		\label{Table 2}
		\begin{tabular}{llllll}
			\hline
			\hline 
			Author & ${\rm d[Fe/H]/d}R$ & ${\rm d[Fe/H]/d}z$ & ${\rm d[\alpha/Fe]/d}z$ &  ${\rm dV_{\phi}/d[Fe/H]}$ & Notes \\
			
			&  (dex ${\rm kpc^{-1}}$)  &  (dex ${\rm kpc^{-1}}$) &  (dex ${\rm kpc^{-1}}$)  &   (${\rm km\ s^{-1}\ dex^{-1}}$) & 
			\\
			\hline
			\multicolumn{6}{c}{Thick disk}\\
			\hline
			\cite{Chen11} & - & -0.12$\pm$0.01 & -&- & RHB stars, $0.5<|z|<3$ kpc\\
			& - & -0.22$\pm$0.07 & -&- & RHB stars, $1<|z|<3$ kpc\\
			
			\cite{Katz11} & - & -0.068$\pm$0.009 & -&- & -\\
			
			\cite{Lee11} & - & - & -&+45.8 $\pm$ 2.9 & G-type dwarfs\\
			
			\cite{Bilir12} & +0.017 $\pm$ 0.008 & -0.034 $\pm$ 0.003 & -&-  & Red clump stars,TD/D$>$10\\
			
			\cite{Coskunoglu12} & $+0.016 \pm 0.011$ & - & -& - &  F-type dwarfs, TD/D$>$10\\
			& $+0.010 \pm 0.009$ & - & -&- & G-type dwarfs, TD/D$>$10\\
			
			\cite{Adibekyan13} & - & - & -&+41.9 $\pm$ 18.1 & FGK-type dwarfs\\
			
			\cite{Mikolaitis14}  & +0.008 $\pm$ 0.007 & -0.072 $\pm$ 0.006 & +0.033 $\pm$ 0.002 & -  & Main \\
			& -0.021  $\pm$ 0.029 & -0.037 $\pm$ 0.016 & +0.011$\pm$ 0.005 & -  & Clean\\
			
			\cite{Recio-Blanco14} & $+0.006 \pm 0.008$ & - & - &+43$\pm$13 & FGK-type stars\\
			
			\cite{Guiglion15}  & - & - & -&+49$\pm$10& FGK-type stars\\
			
			\cite{Prieto16}  & - & - & -&+23 $\pm$ 10 & -\\
			
			\cite{Li17} & $+0.035 \pm 0.01$ & -0.164$\pm$0.010 & -&- & Giants\\
			
			\cite{Duong18}  &- & -0.058$\pm$0.003 &+0.007$\pm$0.002&- & High-$\alpha$ population\\
			
			\cite{Li18} & $+0.031 \pm 0.001$ & -0.086$\pm$0.001 & -0.001$\pm$0.001&- &-\\
			
			\cite{Peng18} & -0.001$\pm$0.020 & - & -&+41.7 $\pm$ 6.1 & GK-type dwarfs\\
			
		    \cite{Guctekin19} & - &  -0.164$\pm$0.014 &-&- & $6<R<10$ kpc and $2<|z|<5$ kpc \\
			
			This work & -0.0006$\pm$0.0005 &  -0.074$\pm$0.0009 &+0.008$\pm0.0002$&+30.87 $\pm$ 0.001 & AFGK-type giants\\

			\hline	
			\multicolumn{6}{c}{Thin disk}\\
			\hline	
			\cite{Lee11} & - & - & -&-22.6 $\pm$ 1.6 & G-type dwarfs\\
			
			\cite{Bilir12} & -0.041 $\pm$ 0.003 & -0.109 $\pm$ 0.008 & -&- & Red clump stars,TD/D$\le$0.1\\
			
			\cite{Coskunoglu12} & $-0.043 \pm 0.005$ & - & -& - & F-type dwarfs, TD/D$\le$0.1\\
			& $-0.033 \pm 0.007$ & - & -& - & G-type dwarfs, TD/D$\le$0.1\\
			
			\cite{Adibekyan13} & - & - & -&-16.8 $\pm$ 3.7 & FGK-type dwarfs\\
			
			\cite{Mikolaitis14}  & -0.044 $\pm$ 0.009 & --0.107$\pm$0.009& +0.041$\pm$0.004 &  -  & Main \\
			& -0.028  $\pm$ 0.018 & -0.057$\pm$0.016 & +0.036$\pm$0.006 &-   & Clean\\
			
			\cite{Recio-Blanco14}  & +0.058 $\pm$ 0.008 & - & - & -17$\pm$6 & FGK-type stars \\
			
			\cite{Guiglion15}  & - & - & -&+4$\pm$3& FGK-type stars\\
			
			\cite{Prieto16}  & - & - & -&-18 $\pm$ 2 & -\\
			
			\cite{Duong18}  &- & -0.18$\pm$0.01 &+0.008$\pm$0.002&-& Low-$\alpha$ population\\
			
			\cite{Li18} & $-0.044 \pm 0.001$ & -0.091$\pm$0.001 & +0.022$\pm$0.001&- & -\\
			\cite{Peng18} & -0.027$\pm$0.031 & - & -&-18.2 $\pm$ 2.3 & GK-type dwarfs\\
			
			\cite{Guctekin19} & -0.042$\pm$0.011 &  -0.308$\pm$0.018 &-&- & F-G type main-sequence stars\\
			
			This work & -0.05$\pm$0.0002 & -0.12$\pm$0.0007 & +0.05$\pm$0.0002& -17.03 $\pm$ 0.001 & AFGK-type giants\\
			\hline
		\end{tabular}
	\end{center}
\end{table*}

\subsection{Vertical chemical abundance gradients of the Galactic disk}

\begin{figure}[]
	\includegraphics[width=1.0\hsize]{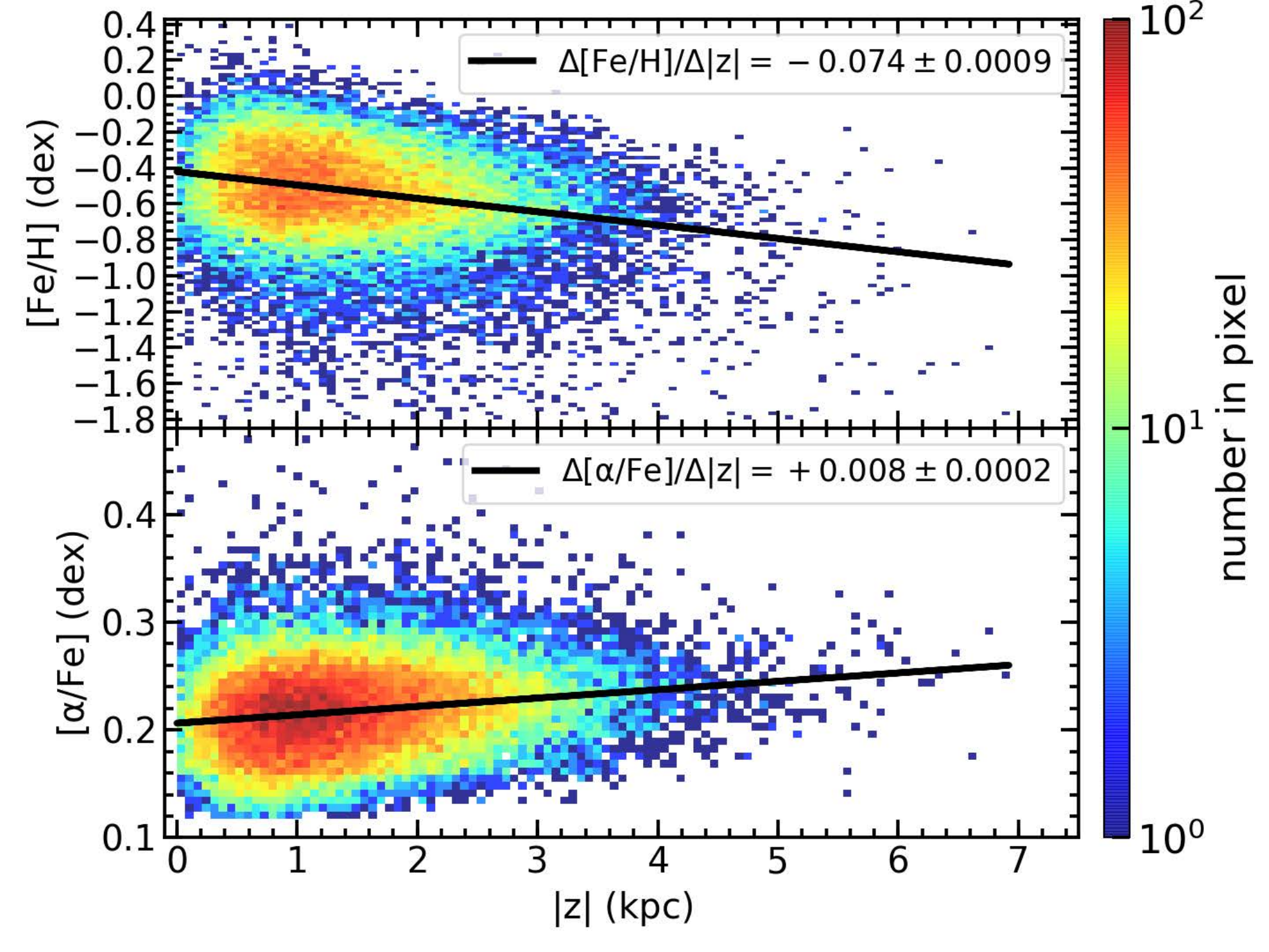}
	\caption{Variation of metallicity and   $\alpha$-abundance with vertical height for the thick disk stars. Top panel:  $\alpha$-abundance gradient with $|z|$ for the thick disk stars. Bottom panel: metallicity gradient with $|z|$ for the thick disk stars.}
	\label{figure7}
\end{figure}

\par 
We also consider the variations of metallicity [Fe/H] and $\alpha$-abundance distributions versus vertical distance $|z|$ for the disk stars and study their vertical gradient. The metallicity  vertical gradient of the thick disk is ${\rm d[Fe/H]/d}|z| = -0.074\pm0.0009$ dex ${\rm kpc^{-1}}$ as shown in the top panel of Figure \ref{figure7}. This result is consistent roughly with the results from some previous works, such as \cite{Katz11}, \cite{Mikolaitis14}, \cite{Duong18} in Table \ref{Table 2}. But \cite{Chen11} used a sample of 1728 red horizontal-branch (RHB) stars with $0.5<|z| < 3$ kpc from SDSS DR8 and \cite{Li17} used 2035 thick-disk giant stars from LAMOST DR3 to derive a steeper gradient of the thick disk  than our result.

\par 
The bottom panel of Figure \ref{figure7} shows that  $\alpha$-abundance gradient with $|z|$ for the thick disk, which derives a very flat gradient of ${\rm d[\alpha/Fe]/d}|z| = +0.008\pm0.0002$ dex ${\rm kpc^{-1}}$. \cite{Duong18} and  \cite{Li18} also found an almost flat gradient  for the thick disk, which are in good agreement with our result. \cite{Mikolaitis14}, using about 2000 F/G/K-type dwarfs and giants from the GES DR1, measured a steeper gradient for the thick disk, ${\rm d[\alpha/Fe]/d}|z| = +0.033 \pm 0.002$ dex ${\rm kpc^{-1}}$ for the main sample. When they use a clean sample (S/N $ > 40$), the result is ${\rm d[\alpha/Fe]/d}|z| = +0.011 \pm 0.005$ dex ${\rm kpc^{-1}}$ which is consistent with our result.

\par 
\begin{figure}
\includegraphics[width=1.0\hsize]{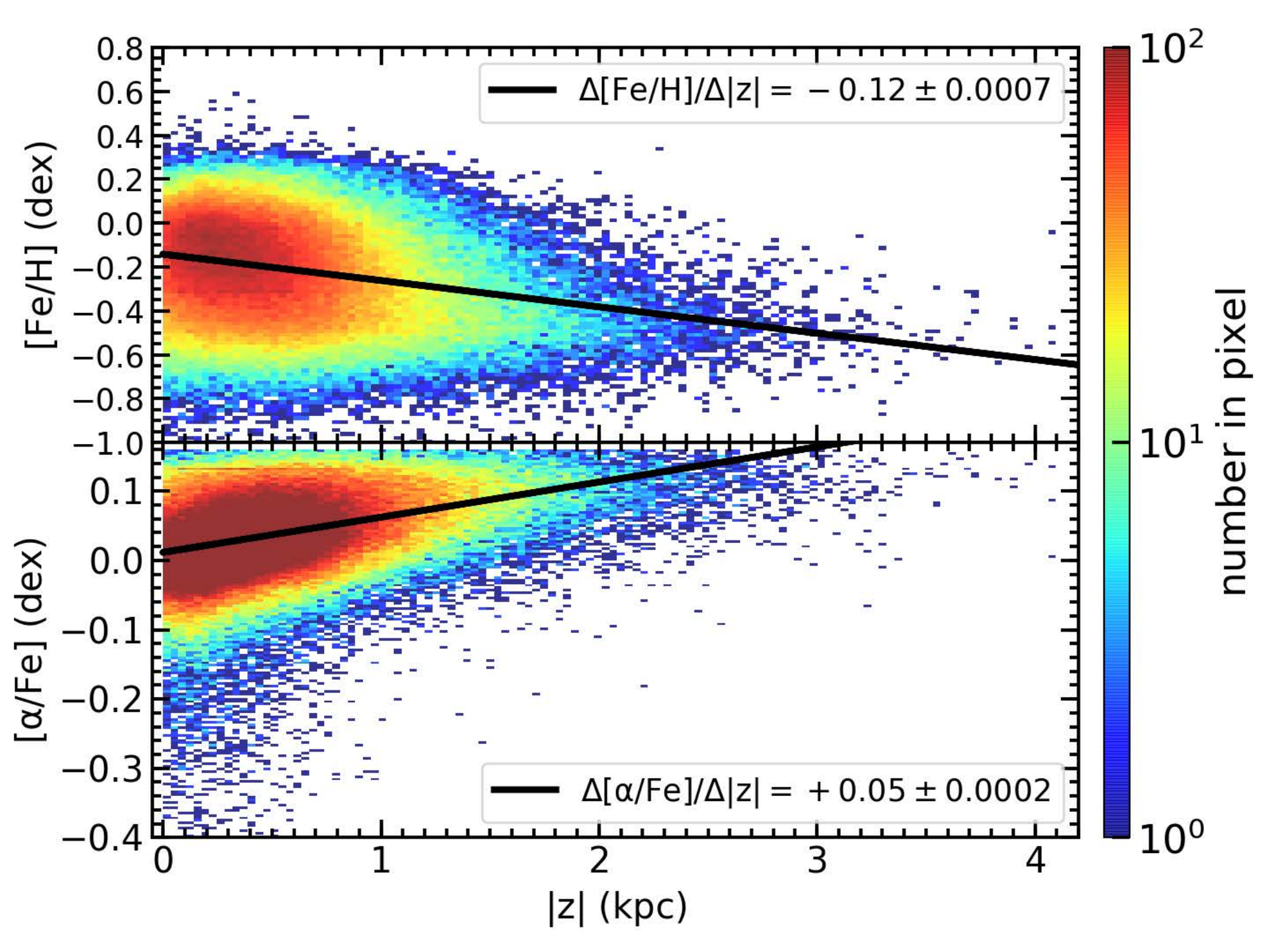}
\caption{Variation of metallicity and $\alpha$-abundance with vertical height for the thin disk. Top panel: metallicity gradient with $|z|$ for the thin disk. Bottom panel: [$\alpha$/Fe] gradient with $|z|$ for the thin disk.}
\label{figure8}
\end{figure}

Figure \ref{figure8} also shows that the trends of vertical metallicity and  $\alpha$-abundance for the thin disk. The top panel of this figure indicates the thin disk has a obvious metallicity  vertical gradient ${\rm d[Fe/H]/d}|z| = -0.12\pm0.0007$ dex ${\rm kpc^{-1}}$, which is in good agreement with the results from previous works. For example, \cite{Bilir12}, \cite{Duong18}, \cite{Li18} reported ${\rm d[Fe/H]/d}|z|=-0.109\pm0.008$ dex ${\rm kpc^{-1}}$, ${\rm d[Fe/H]/d}|z|=-0.18\pm0.01$dex ${\rm kpc^{-1}}$ and ${\rm d[Fe/H]/d}|z| = -0.09 \pm 0.001$dex ${\rm kpc^{-1}}$, respectively. 
\par 
The [$\alpha$/Fe] gradient with vertical height for the thin disk stars is given in the bottom panel of Figure \ref{figure8}. The [$\alpha$/Fe]  vertical gradient of the thin disk is ${\rm d[\alpha/Fe]/d}|z|= +0.05\pm0.0002$ dex ${\rm kpc^{-1}}$. This result agrees with the result of \cite{Mikolaitis14}  ${\rm d[\alpha/Fe]/d}|z|= +0.041 \pm 0.004$ dex ${\rm kpc^{-1}}$, and is slightly steeper than result given by \cite{Li18} ${\rm d[\alpha/Fe]/d}|z|= +0.022 \pm 0.0001$ dex ${\rm kpc^{-1}}$. However, \cite{Duong18} derived a very flat gradient ${\rm d[\alpha/Fe]/d}|z|= +0.008 \pm 0.002$ dex ${\rm kpc^{-1}}$ using data from the GALAH survey internal data release.

\subsection{Variation of metallicity gradients with radial distance and vertical height}
\par
\begin{figure}[]
	\includegraphics[width=1.0\hsize]{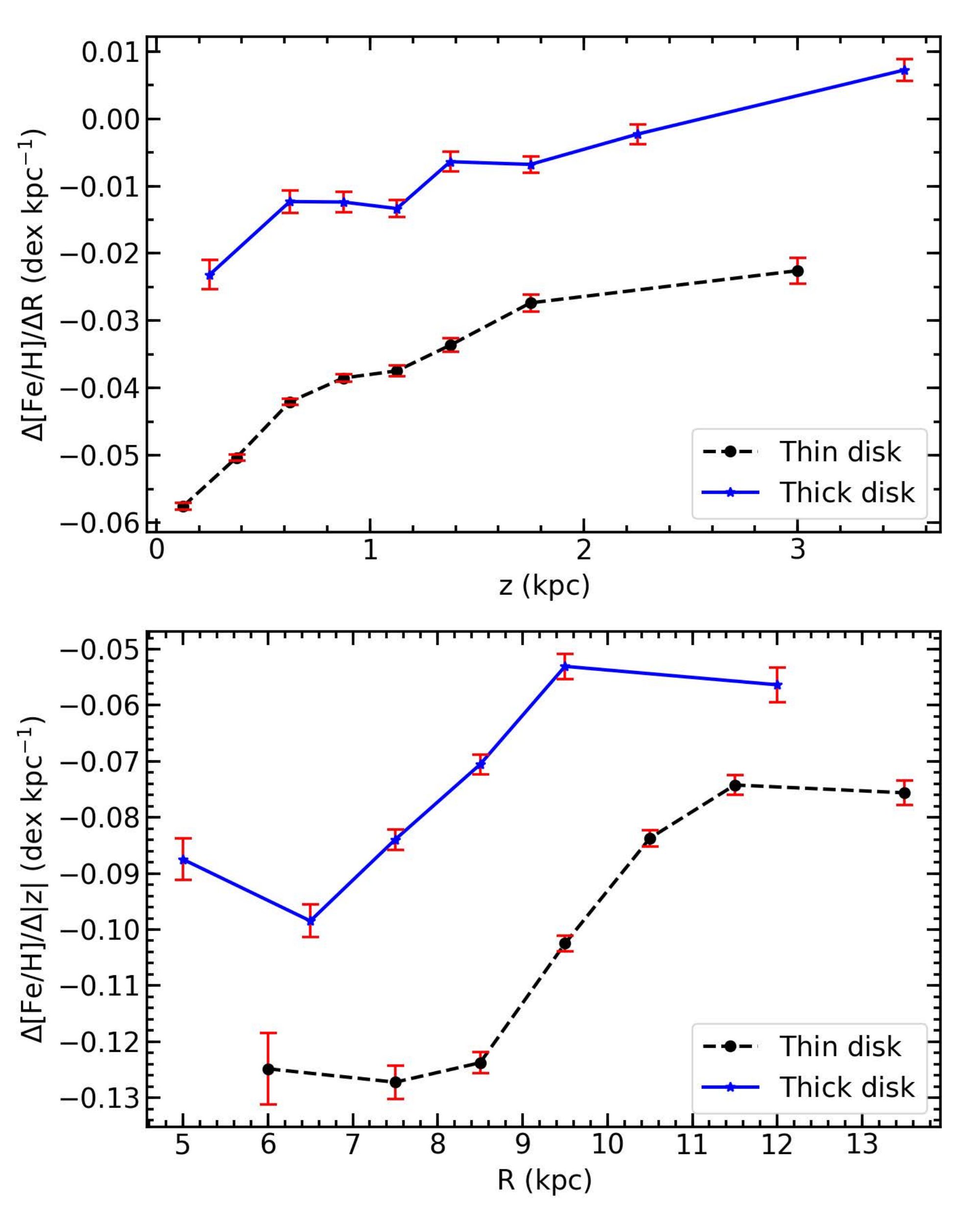}
	\caption{Top panel: variation of metallicity radial gradients versus vertical distance for the thin disk (black dotted line) and thick disk (blue solid line). Bottom panel: the variation of metallicity vertical gradient versus radial distance for the thin disk (black dotted line) and thick disk (blue solid line). Points of data are from Table \ref{Table 3} for the top panel and Table \ref{Table 4} for the bottom panel.}
	\label{figure9}
\end{figure}

 We now study the variation of metallicity radial  gradients versus vertical height and metallicity vertical gradients versus radial distance for the thin and thick disk stars. Several subsamples are separated in Galactocentric radial distance and vertical height. For each subsample, the gradient and its error are obtained by the method introduced in section 3.1. Results of gradients and the number of stars in each subsample are given in Table \ref{Table 3} and Table \ref{Table 4}.  The curves in Figure \ref{figure9} are simple connection of data points from Table \ref{Table 3} or Table \ref{Table 4}.
\par 

In Figure 9, the top panel gives the variation of  metallicity radial gradient versus vertical distance and it shows that it flattens with increasing vertical height for the thin disk, and the thick disk (blue solid line) also flattens slightly with increasing vertical height. The bottom panel gives the variation of metallicity  vertical gradient versus radial distance, which also shows almost constant in the inner region and  a flattening with increasing radial distance in the outer region for the thin disk, while the thick disk (blue dashed line) is always slightly flat on average.

\section{Kinematics Properties of the Galactic Disk}
\subsection{The distribution of stellar orbital eccentricities}

\begin{figure}[]
	\includegraphics[width=1.0\hsize]{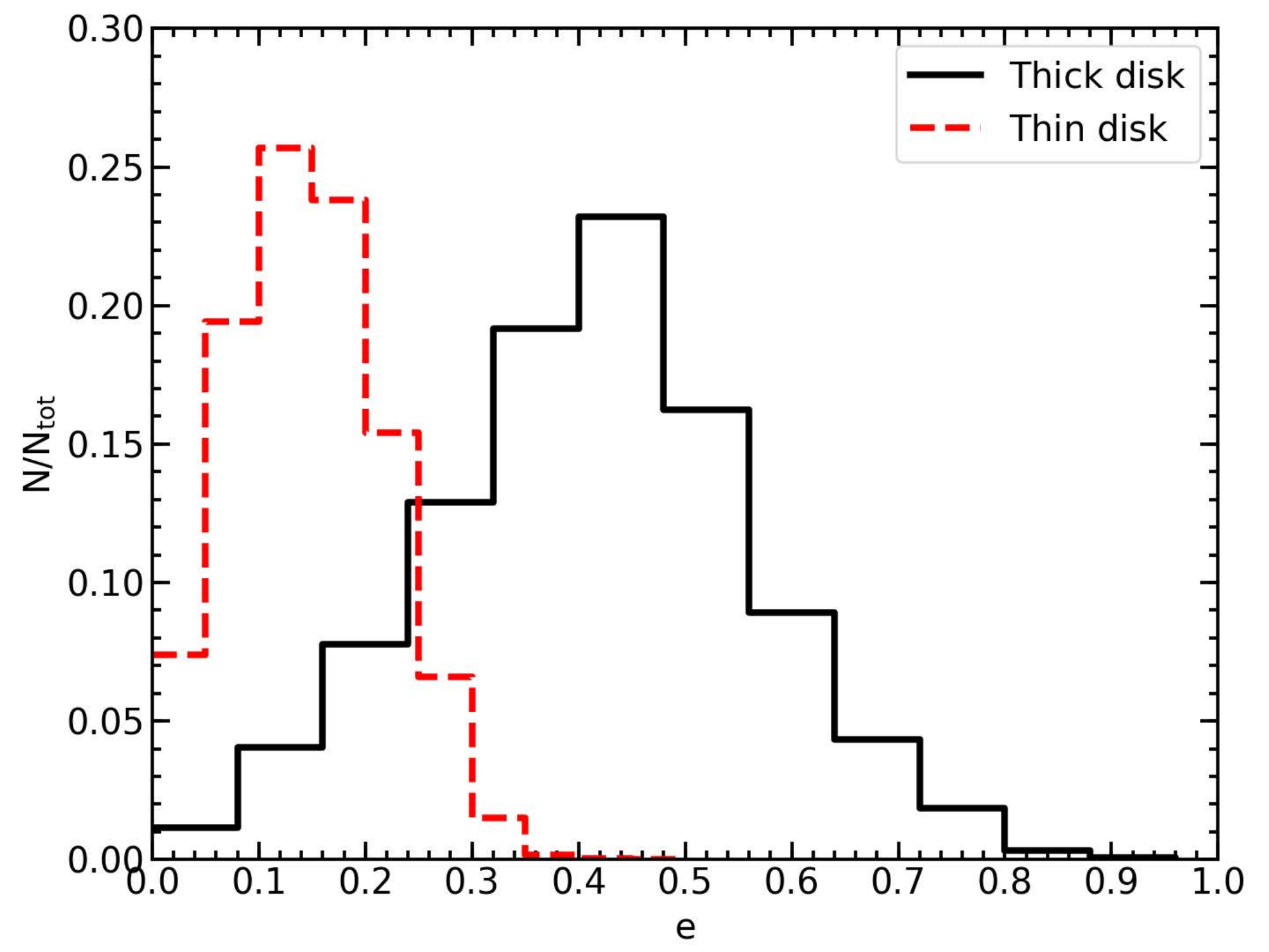}
	\caption{Top panel: normalized distributions of orbital eccentricities for the thick disk (black dotted line) and thin disk (red solid line).}
	\label{figure10}
\end{figure}
\par 
We study the orbital properties of the thin and thick disk stars by adopting a Galaxy potential model. We use recent Galactic potential model provided by \cite{McMillan17}. This model includes components that represent the contribution of the cold gas discs near the Galactic plane, as well as the thin and thick stellar disk, a bulge component and a dark-matter halo. Based on the this Galactic potential model,  we compute the orbital eccentricities of the thin and thick disk stars, $e$, defined as $e= (r_{\rm apo} - r_{\rm peri})/(r_{\rm apo} + r_{\rm peri})$, where $ r_{\rm peri}$ and $r_{\rm apo}$ denote the closest approach of an orbit to the Galactic center and the farthest extent of an orbit from the Galactic center. 

\par 
Figure \ref{figure10} shows the normalized distributions of orbital eccentricities for the thick and thin disk stars. It can be seen that the distribution of orbital eccentricities for the thick disk and thin disk stars have obviously  different distribution. The eccentricities distribution of the thin disk sample stars peaks at $e\sim 0.12$, with narrow widths, and includes very few high eccentricities stars ($e > 0.3$). In contrast, the distribution for the thick disk stars shows higher eccentricities and peaks at $e\sim 0.42$, with wide widths, and extend higher eccentricities up to $e \sim 0.8$. 
\par 
\cite{Sales09} demonstrated that the orbital eccentricities distribution could help to probe the formation mechanisms of the thick disk.  Four simulation model for the thick disk formation are provided:  radial migration model \citep[e.g.,][]{Sellwood02, Binney09}, gas-rich merger models \citep[e.g.,][]{Brook04,Brook05}, the accretion model \citep{Abadi03} and Heating scenario \citep[e.g.,][]{Quinn93, Kazantzidis08}. 
Generally, the distribution of stellar orbital eccentricities generated by violent models such as disk heating and accretion include higher eccentricity stars ($e > 0.6$). The discussion of the formation mechanisms of the thick disk is given in section 5.

\begin{table}[hbp]
	\begin{center}
		\centering
		\caption{ \upshape {The number, metallicity gradients with radial distance and mean orbital eccentricities (and its standard deviation) of the stars at different vertical bins for the thick and thin disk. } }
		\label{Table 3}
		\begin{tabular}{lllll}
			\hline
			\hline 
			Vertical height & $N_{\rm stars}$ & ${\rm d[Fe/H]/d}R$ & Mean $e$ & $\sigma_e$\\
			
			(kpc)    &                                &          (dex ${\rm kpc^{-1}}$)    &    &    
			\\
			\hline
			\multicolumn{5}{c}{Thick disk}\\
			\hline
			$0\le|z|<0.5$ & 3233 & $-0.0231\pm 0.002$ & 0.4123 & 0.137\\
			$0.5\le|z|<0.75$& 3416& $-0.0123\pm 0.0017$ & 0.4008 & 0.144\\
			$0.75\le|z|<1$& 3992& $-0.0124\pm 0.0016$ & 0.3938 & 0.145\\
			$1\le|z|<1.25$& 3926& $-0.0133\pm 0.0012$& 0.3959 & 0.148\\
			$1.25\le|z|<1.5$& 3459& $-0.0062\pm 0.0013$ & 0.401 & 0.148\\
			$1.5\le|z|<2$& 5114& $-0.0068\pm 0.0012$ & 0.4161 & 0.147\\
			$2\le|z|<2.5$& 3294& $+0.0023\pm 0.0013$ & 0.4300 & 0.155\\
			$2.5\le|z|<4.5$& 3277& $+0.0073\pm 0.0014$ & 0.47 & 0.154\\
			\hline	
			\multicolumn{5}{c}{Thin disk}\\
			\hline	
			$0\le|z|<0.25$ & 48514 & $-0.0576\pm 0.0005$& 0.1461 & 0.066\\
			$0.25\le|z|<0.5$& 51968& $-0.0503\pm 0.0005$& 0.1459 & 0.069\\
			$0.5\le|z|<0.75$& 38328& $-0.0421\pm 0.0004$& 0.1502 & 0.070\\
			$0.75\le|z|<1$& 20157& $-0.0385\pm 0.0005$& 0.1561 & 0.071\\
			$1\le|z|<1.25$& 9802& $-0.0374\pm 0.0008$ & 0.1549 & 0.073\\
			$1.25\le|z|<1.5$& 4983& $-0.0336\pm 0.001$& 0.1507 & 0.073\\
			$1.5\le|z|<2$& 3858& $-0.0274\pm 0.001$&0.1467 & 0.074\\
			$2\le|z|<4$& 1476& $-0.0226\pm 0.002$ & 0.1420& 0.072\\
			\hline
		\end{tabular}
	\end{center}
\end{table}
\subsection{The distribution of stellar orbital eccentricities with metallicity and vertical height}

\begin{table}[hbp]
	\begin{center}
		\centering
		\caption{ \upshape {The number and metallicity gradients with vertical height of the stars at different radial bins for the thick and thin disk. } }
		\label{Table 4}
		\begin{tabular}{lll}
			\hline
			\hline 
			Radial distance & $N_{\rm stars}$ & ${\rm d[Fe/H]/d}|z|$ \\
			
			(kpc)    &                                &          (dex ${\rm kpc^{-1}}$)       
			\\
			\hline
			\multicolumn{3}{c}{Thick disk}\\
			\hline
			$4\le R<6$ & 2045 & $-0.0874\pm 0.0036$\\
			$6\le R<7$& 3434& $-0.098\pm 0.0026$\\
			$7\le R<8$& 7820& $-0.0839\pm 0.0018$\\
			$8\le R<9$& 8187& $-0.0705\pm 0.0017$\\
			$9\le R<10$& 4692& $-0.053\pm 0.0024$\\
			$10\le R<14$& 3670& $-0.056\pm 0.0031$\\
			\hline	
			\multicolumn{3}{c}{Thin disk}\\
			\hline	
			$5\le R<7$ & 1565 & $-0.1252\pm 0.0069$\\
			$7\le R<8$& 11907& $-0.1272\pm 0.0030$\\
			$8\le R<9$& 35712& $-0.1237\pm 0.0017$\\
			$9\le R<10$& 57326& $-0.1025\pm 0.0015$\\
			$10\le R<11$& 36618& $-0.0837\pm 0.0016$\\
			$11\le R<12$& 22360& $-0.0742\pm 0.0019$\\
			$12\le R<15$& 13567& $-0.0756\pm 0.0024$\\
			\hline
		\end{tabular}
	\end{center}
\end{table}
\par
\begin{figure}[]
	\includegraphics[width=1.0\hsize]{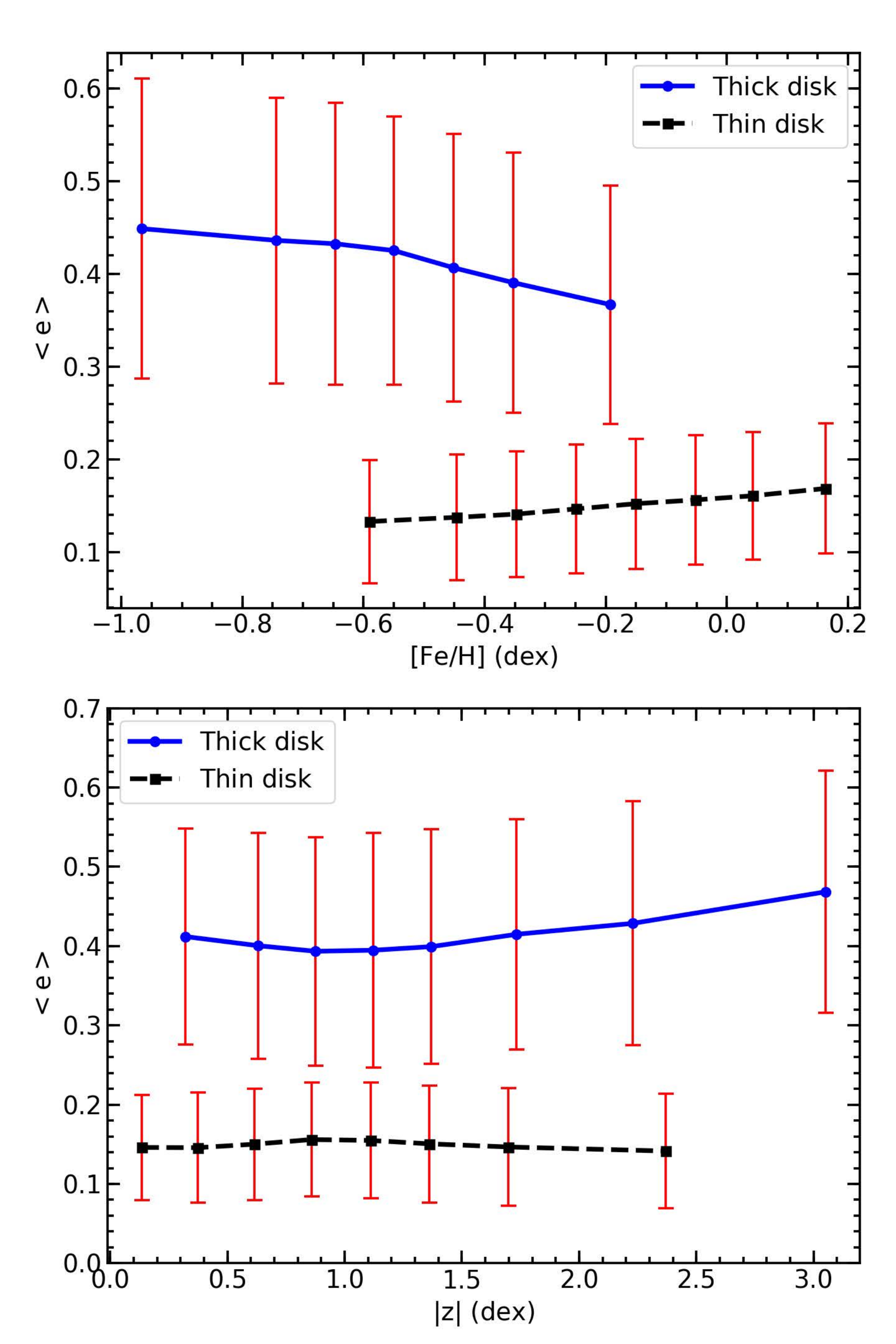}
	\caption{The variation of orbital eccentricity with metallicity  (top panel) and vertical height (bottom panel) for the thick disk and thin disk (bottom panel). The error bar stands for standard deviation in each subsample. Points of data are from Table \ref{Table 5} for the top panel and Table \ref{Table 3} for the bottom panel.}
	\label{figure11}
\end{figure}

 
We study the variation of  orbit eccentricities with metallicity and vertical height for the thin and thick disk. Several subsamples are separated in metallicity and vertical height.  More detailed information about subsample is listed in Table \ref{Table 3} and \ref{Table 5}. The trend of orbital eccentricities with metallicity and vertical height is shown in Figure \ref{figure11}, and the curves are the simple connection of data points from Table \ref{Table 3} or Table \ref{Table 5}.
\par 
The top panel of Figure \ref{figure11} indicates that average orbital eccentricities exist a downtrend with increasing metallicity, but it is almost constant in the range of $-1.0 \lesssim {\rm [Fe/H]} \lesssim -0.6$ dex for the thick disk. In general, the stars with poor metallicity form in an earlier time, which means that the thick disk stars which have lower orbit eccentricity are the youngest ones. We could think that it is possible that these young stars in the thick disk could originate in the thin disk, reaching the thick disk through radial migration or other mechanisms.  However, the orbital eccentricity has a slight uptrend with increasing metallicity for the thin disk. It also indicates that stars having the highest orbit eccentricities are probably  the youngest since they are the most metal-rich. These two trends of orbital eccentricities are consistent with the results of \cite{Lee11} and \cite{Peng18}. The bottom panel of Figure \ref{figure11} indicates that there is no obvious correlation between orbital eccentricities and vertical height for the thick disk and the thin disk.
\par 
We discuss the impact of potentially mixing between the thin disk and thick disk in $-0.6<$[Fe/H]$<-0.2$ dex on trends of the top panel of Figure \ref{figure11}. If our thick disk sample with high eccentricity has a strong effect on the thin disk in the top panel of Figure \ref{figure11}, average orbit eccentricity in $-0.6<$[Fe/H]$<-0.2$ dex should be higher than relative rich thin disk stars. However, the  orbital eccentricity of the thin disk has a uptrend with increasing metallicity. Furthermore, the number of the thin disk stars is six times that of thick disk stars in our samples. Thus, the thick disk has little effect on the thin disk. The top panel of Figure \ref{figure11} indicates that average orbital eccentricity of the thick disk has a downtrend with increasing metallicity within $-0.6<$[Fe/H]$<-0.2$ dex. In order to confirm this  downtrend is not affected by the thin disk stars with low eccentricity, we strengthen the conditions for the selection of thick disk stars within $-0.6<$[Fe/H]$<-0.2$, and we restrict TD/D $> 20$ (TD/D has been explained in section 2.3) and $e>0.2$ for the thick disk stars in $-0.6<$[Fe/H]$<-0.2$. We find that the orbital eccentricities trend  for restricted thick disk stars  is similar to the top panel of Figure \ref{figure11}. Thus, it indicates that the thin disk has little effect on the thick disk in Figure \ref{figure11}. It also indicates potential mixing between the thin disk and thick disk has little effect on Figure \ref{figure10}.
\begin{table}[hbp]
	\begin{center} 
		\centering
		\caption{ \upshape {The number and  mean orbital eccentricities (and its standard deviation) of the stars at different [Fe/H] bins for the thick disk and thin disk. } }
		\label{Table 5}
		\begin{tabular}{llll}
			\hline
			\hline 
			[Fe/H] & $N_{\rm stars}$ & Mean $e$ & $\sigma_e$ \\
			
			(dex)    &               &                 &                
			\\
			\hline
			\multicolumn{4}{c}{Thick disk}\\
			\hline
			$-1.8\le {\rm [Fe/H]}<-1.4$ & 266 & - & -\\
			$-1.4\le {\rm [Fe/H]}<-0.8$ & 3775 & $0.4571$ & 0.165\\
			$-0.8\le {\rm [Fe/H]}<-0.7$& 2793& 0.4366 & 0.154\\
			$-0.7\le {\rm [Fe/H]}<-0.6$& 4722& 0.4330 & 0.152\\
			$-0.6\le {\rm [Fe/H]}<-0.5$& 5221& 0.4259 & 0.145\\
			$-0.5\le {\rm [Fe/H]}<-0.4$& 4723& 0.4070 & 0.145\\
			$-0.4\le {\rm [Fe/H]}<-0.3$& 3909& 0.3908 & 0.141\\
			$-0.3\le {\rm [Fe/H]}$& 4823& 0.3660 & 0.129\\
			\hline	
			\multicolumn{4}{c}{Thin disk}\\
			\hline	
			$-1\le {\rm [Fe/H]}<-0.5$ & 15646 & $0.1306$ & 0.066\\
			$-0.5\le {\rm [Fe/H]}<-0.4$& 19450& 0.1377 & 0.068\\
			$-0.4\le {\rm [Fe/H]}<-0.3$& 26090& 0.1413 & 0.068\\
			$-0.3\le {\rm [Fe/H]}<-0.2$& 31150& 0.1467 & 0.069\\
			$-0.2\le {\rm [Fe/H]}<-0.1$& 32297& 0.1522 & 0.072\\
			$-0.1\le {\rm [Fe/H]}<0$& 28962& 0.1563 & 0.069\\
			$0\le {\rm [Fe/H]}<0.1$& 17201& 0.1607 & 0.069\\
			$0.1\le {\rm [Fe/H]}$& 8222& 0.1686 & 0.070\\
			\hline
		\end{tabular}
	\end{center}
\end{table}
\subsection{The rotational velocity gradients with radial distance, vertical height and metallicity}
\par 
The gradients of rotation velocity are important properties for the Galactic disk and can provide a useful clue to the formation and evolution of Galactic disk. For example, the gradients of rotation velocity were predicted by some models of disk formation, and it was observed by some studies. Thus, it provides observational clues for the model prediction.
 Also, it can restrict model parameters, for example, \cite{Curir12} reported that the gradient of $V_{\phi}$ with [Fe/H] of thick disk play a critical role to determine the variation trend of the metallicity of Galactic disk with $R$ at the time of Galactic disk formation. Here, we examine the observed gradients of $V_{\phi}$ with $R$, $|z|$ and [Fe/H] using our sample stars. 

\par
\begin{figure}
	\includegraphics[width=1.0\hsize]{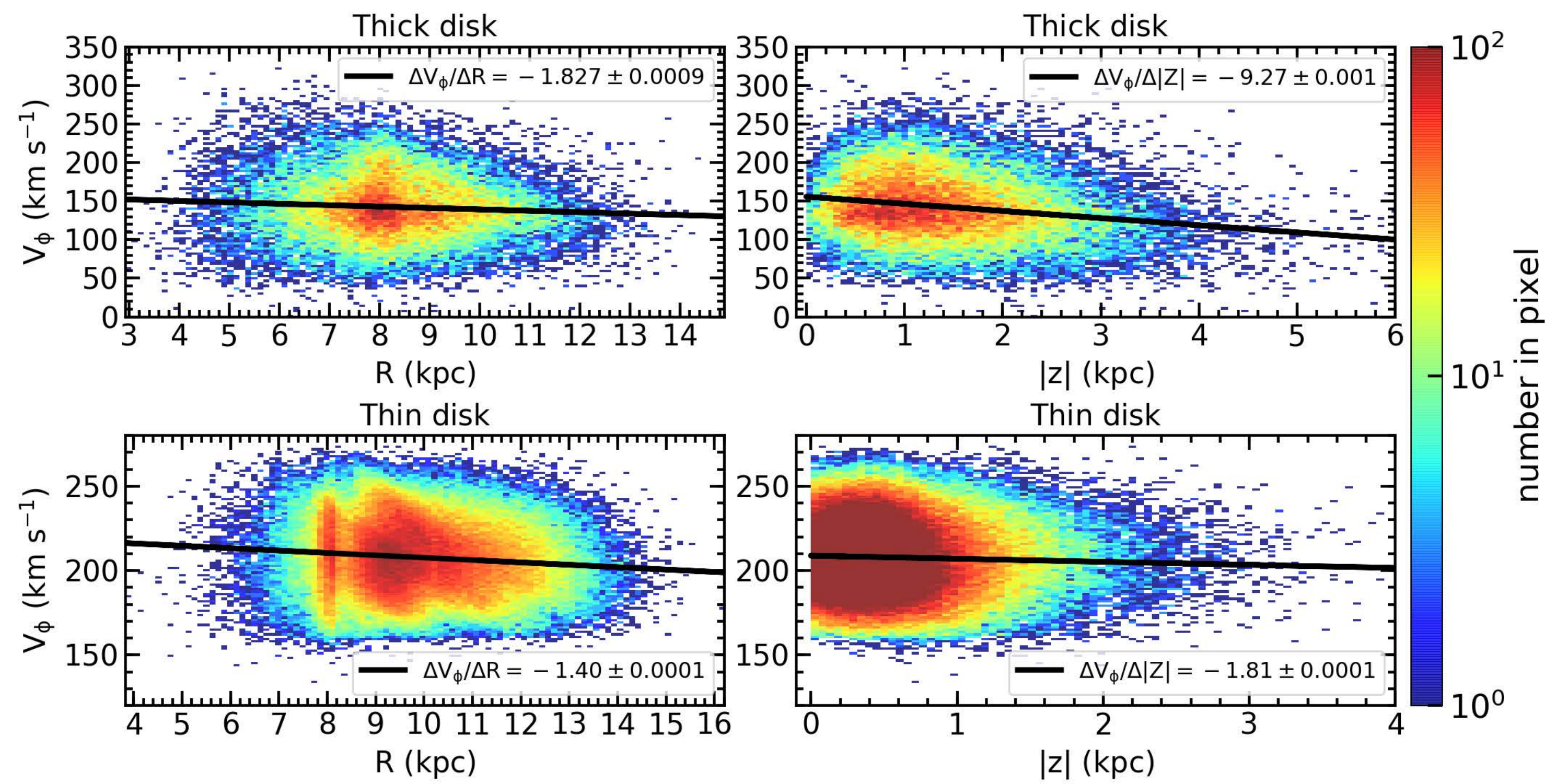}
	\caption{Variation of rotational velocity with radial distance and vertical height for the thin disk and thick disk stars.}
	\label{figure12}
\end{figure}

 Figure \ref{figure12} displays the distribution of rotational velocity with radial distance (left panel) and vertical height (right panel) for the thin disk (bottom panel) and thick disk (top panel). In the radial direction, very small rotational velocity gradients of $-1.827\pm0.0009$  ${\rm km\ s^{-1}\ kpc^{-1}}$ and  $-1.40\pm0.0001$  ${\rm km\ s^{-1}\ kpc^{-1}}$ are derived for the thick disk and the thin disk, respectively.  In the vertical direction, there also exist two small rotational velocity gradients of $-9.27\pm0.001$ ${\rm km\ s^{-1}\ kpc^{-1}}$ and $-1.81\pm0.0001$ ${\rm km\ s^{-1}\ kpc^{-1}}$ for the thick disk and the thin disk, respectively. The results are basically in agreement with \cite{Lee11}. \cite{Lee11} also observed flat gradients: ${\rm dV_{\phi}/d}R = -5.6 \pm 1.1$ ${\rm km\ s^{-1}\ kpc^{-1}}$ and  ${\rm dV_{\phi}/d}z = -9.4 \pm 1.3$ ${\rm km\ s^{-1}\ kpc^{-1}}$ for the thick disk, and ${\rm dV_{\phi}/d}R = -0.1 \pm 0.6$ ${\rm km\ s^{-1}\ kpc^{-1}}$ and ${\rm dV_{\phi}/d}z = -9.2 \pm 1.2$ ${\rm km\ s^{-1}\ kpc^{-1}}$ for the thin disk. 

\par
\begin{figure}[]
	\includegraphics[width=1.0\hsize]{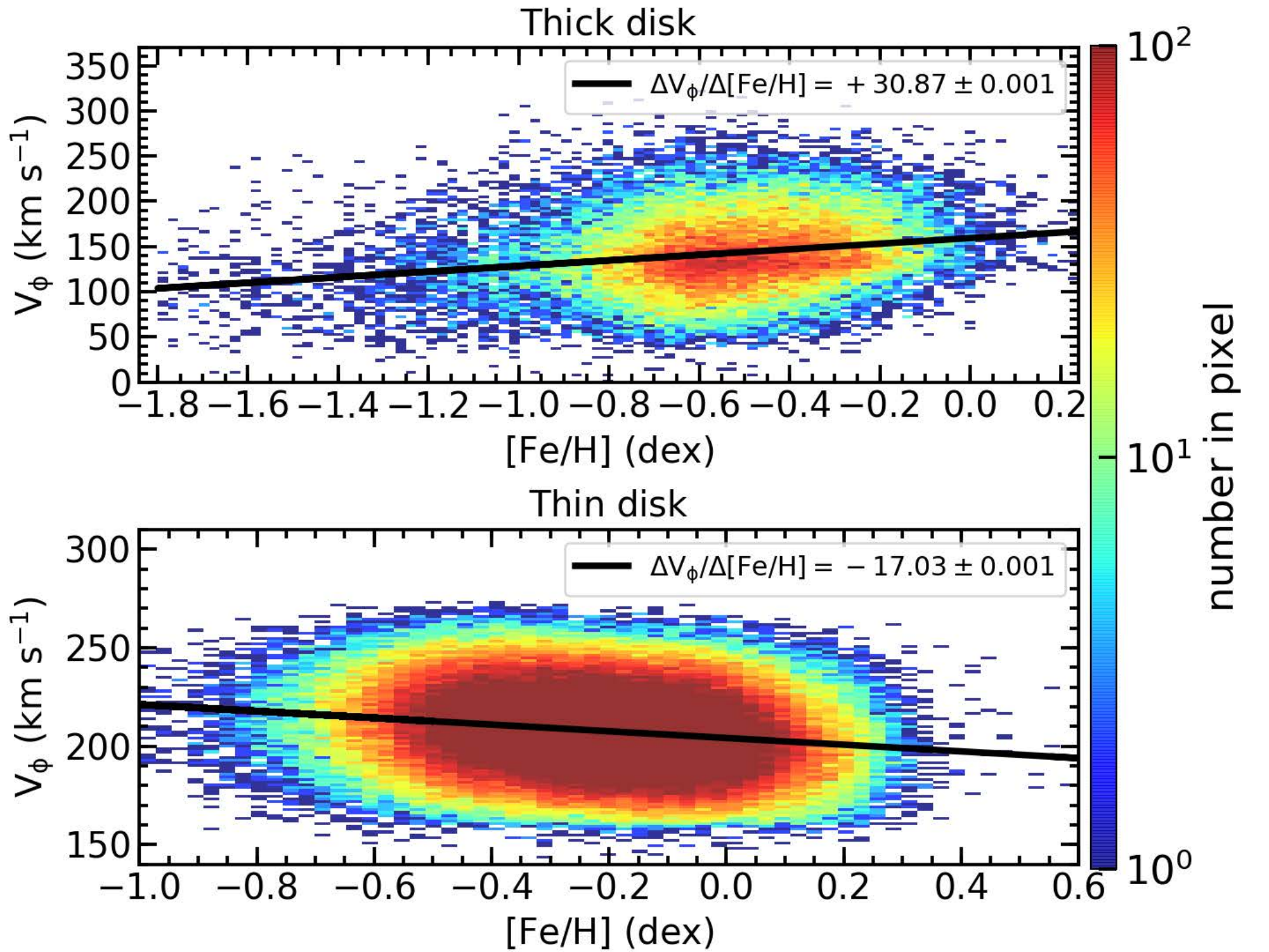}
	\caption{Variation of rotational velocity with metallicity abundance for the thin disk and thick disk stars.}
	\label{figure13}
\end{figure}

Figure \ref{figure13} also indicates that there exist two clear rotational velocity gradients with metallicity, ${\rm dV_{\phi}/d[Fe/H]}=+30.87\pm0.001$ ${\rm km\ s^{-1}\ dex^{-1}}$ for the thick disk  and ${\rm dV_{\phi}/d[Fe/H]} = -17.03\pm0.001$ ${\rm km\ s^{-1}\ dex^{-1}}$ for the thin disk. As shown in Table \ref{Table 2}, many studies also derived a clear rotational velocity gradient with metallicity from -16 to -24 ${\rm km\ s^{-1}\ dex^{-1}}$ for the thin disk,  and from +23 to +46 ${\rm km\ s^{-1}\ dex^{-1}}$ for the thick disk. 
So our results are within the range given by other authors.  

\section{Discussion of Selection Bias and the Formation and Evolution of the Galactic Disk}
\par 

One may be concerned about biases in our sample that might arise from the selection of A/F/G/K-type giant stars, which may lead to loss of  more massive and youngest stars in our sample. Our sample has no massive stars, but this is not a serious problem, because there are considerable low-mass young giant stars according to the result of \cite{Wu19}, who measured the age and mass for a sample of 640, 986 red giant branch (RGB) stars of the Galactic disk from the LAMOST DR4.  Furthermore, the selection of  giant stars may also lead to  bias against old stars and neglect some young stars in our initial sample. Some previous studies suggest that ages are about 9-10 Gyr for the high-[$\alpha$/Fe] stars in the Galactic disk \citep[e.g.,][]{Bensby14,Bergemann14,Wu19}, thus, the thick disk may contain few young stars, and this selective bias can have little effect on the thick disk. If selecting giant sample stars make strong favors old stars over young stars in the thin disk, the metallicity distribution of the thin disk can be shifted to lower. Thus, this bias might produce misleading correlations between [Fe/H] versus the spatial ($R$,$|z|$) and kinematic parameters ($V_{\phi}$). However, as discussed above, our results are well consistent with  other studies for the thin disk. Therefore, the impact of this bias on the thin disk may also be negligible. 
\par 
We discuss the implication of our results for the formation and evolution of the Galactic disk. Because a model or simulation  is affected by many factors, such as unavoidable numerical effects, and some
assumptions that are required in their construction, models and simulations about formation and evolution of the Galactic disk 
may be not a complete physical reality. Thus, it is difficult to compare our observational results with expectations from the results of models and simulations quantitatively. So we only compare qualitatively our observational results with expectations from the published radial migration, gas-rich, merger, accretion, and disk heating models.  More detailed quantitative comparisons  need to construct more physical realistic models and simulations. 

\par 
First of all, we discuss the formation and evolution of the thin disk. According to the radial migration model \citep[e.g.,][]{Sellwood02, Binney09, Minchev10, McMillan17}, in the inner region of the Galactic disk, gas density is higher and chemical abundance is richer than the outer region.  Thus, most stars formed in the inner disk  should be metal-rich, whereas those born in the outer disk are metal-poor.  It indicates that the thin disk may exist a negative metallicity gradient with radial distance in earlier times. These stars could be affected by radial migration: stars of the thin disk that born in the outer disk move inward to the solar neighborhood, while metal-rich stars formed in the inner disk migrate outward into the solar neighborhood. So the radial migration can flatten the metallicity  radial gradient. \cite{Loebman11} confirmed, using N-body simulations of radial migration, a weak metallicity  radial gradient for the thin disk and shown that the thin disk in solar neighborhood ($R=7-11$ kpc and $|z|=0.3-2.0$ kpc) has a gradient of  ${\rm d[Fe/H]/d}R= -0.02$ dex ${\rm kpc^{-1}}$.  This gradient is slight flatter than our results ${\rm d[Fe/H]/d}R= -0.05 \pm 0.0002$ dex ${\rm kpc^{-1}}$. Furthermore, a rotation velocity gradient of ${\rm dV_{\phi}/d[Fe/H]} = -19$ ${\rm km\ s^{-1}\ dex^{-1}}$
for the thin disk also is found by  \cite{Loebman11} for younger stars (identified with the thin-disk component with low [$\alpha$/Fe]) in solar neighborhood ($R=7-9$ kpc and $|z|=0.1-1$ kpc), which is consistent with our result of ${\rm dV_{\phi}/d[Fe/H]} = -17.03 \pm 0.001$ ${\rm km\ s^{-1}\ dex^{-1}}$. Thus, we conclude that the metallicity  radial gradient and rotation velocity gradient with [Fe/H] for the thin disk can be explained by radial migration model.
\par 
\cite{Brook07}  predicted a correlation between rotation velocity and radial distance for the Galactic disk stars using  N-body simulations of the gas-rich merger model. They suggested there exists a rotation velocity gradient with radial distance for the thin disk (that they refer to as “disk stars”) in the solar neighborhood ($6<R<10$) and there is no [$\alpha$/Fe] gradient with vertical height for the thin disk. These two properties differ from our results of  ${\rm dV_{\phi}/d}R = -1.404\pm0.0001{\rm km\ s^{-1}\ kpc^{-1}}$ and ${\rm d[\alpha/Fe]/d}z= +0.05 \pm 0.0002$ for the thin disk. Thus, we conclude that the gas-rich merger model may not explain the lack of rotation velocity gradient with radial distance and the existence of [$\alpha$/Fe] gradient in vertical direction for the thin disk.
\par 
\begin{figure}[]
	\includegraphics[width=1.0\hsize]{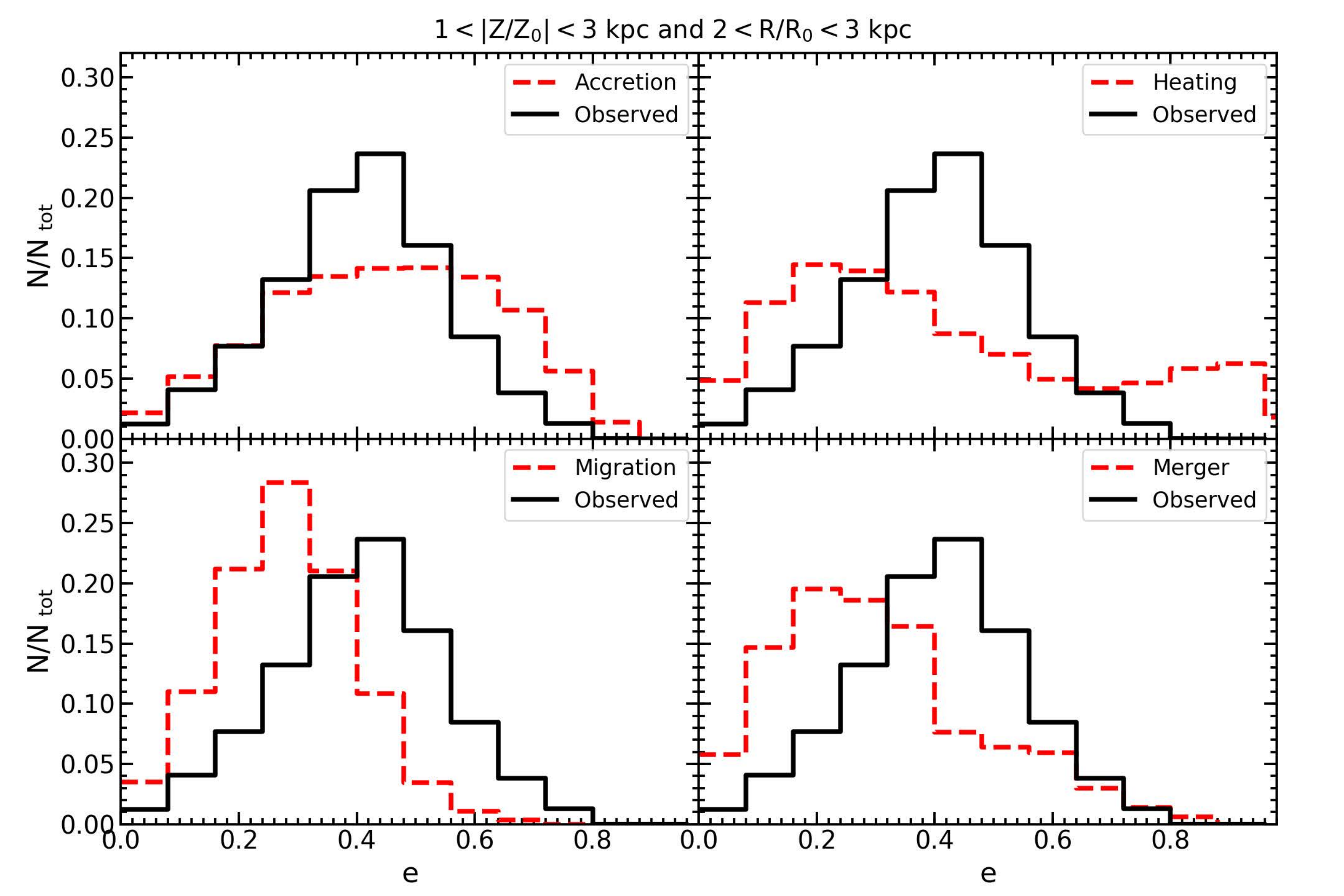}
	\caption{The comparisons of our observed orbital eccentricity distributions for the thick disk with models of disk formation from Figure 3 of \cite{Sales09}. Adopting the scale height and scale length are $z_0=0.8$ kpc and $R_0=3.5$ kpc, which is basically  consistent with Figure 3 of \cite{Sales09}. }
	\label{figure14}
\end{figure}
About the formation and evolution of the thick disk, \cite{Loebman11} reported  a  metallicity radial gradient of  ${\rm d[Fe/H]/d}R = 0.00$ dex ${\rm kpc^{-1}}$ for the thick disk from their N-body simulations of radial migration, and \cite{Loebman16} also shown metallicity vertical gradient of ${\rm d[Fe/H]/d}|z| \approx -0.03$ dex ${\rm kpc^{-1}}$ . These two results are consistent with our results of ${\rm d[Fe/H]/d}R = -0.0006\pm0.0005$ dex ${\rm kpc^{-1}}$ and  ${\rm d[Fe/H]/d}|z| = -0.074\pm0.0009$ dex ${\rm kpc^{-1}}$. However, \cite{Loebman11} reported a rotation velocity gradient of  ${\rm dV_{\phi}/d[Fe/H]} = +8$ ${\rm km\ s^{-1}\ dex^{-1}}$ for older stars (8 $\le$ Age $\le$ 10 Gyr, identified with the thick disk component with high [$\alpha$/Fe]) in solar neighborhood ($R=7-9$ kpc and $|z|=0.1-1$ kpc), which differ from our result of  ${\rm dV_{\phi}/d[Fe/H]} = +30.87 \pm 0.001$ ${\rm km\ s^{-1}\ dex^{-1}}$. Figure \ref{figure14} gives comparisons of observed orbital eccentricities with the accretion, heating, radial migration and merger model depicted the formation scenarios of the thick disk from \cite{Sales09}. Our result shows that the radial migration model lack high orbital eccentricities stars. 
In summary, radial migration model can explain metallicity radial and vertical gradients but not rotation velocity gradients and distribution of orbital eccentricities for the thick disk.
Maybe it indicates that radial migration could not have played a major role in the formation and evolution of the thick disk, it's just involved in the process of formation and evolution of the thick disk.
\par 
According to the N-body simulations model of the dynamical heating of a pre-existing thin disk, \cite{Villalobos10} reported that the $V_{\phi}$ gradients with $R$ and $|z|$ depend sensitively on the orbital inclination of the infalling satellite that produced the initial thick disk. They shown that the thick disk exhibits a very weak trend of $V_{\phi}$ with $R$ for the low orbital inclination, and the correlation between the two quantities becomes stronger with the incidence angle increased. However, the correlation between $V_{\phi}$ and $|z|$ exhibits very weak for the high  orbital inclination and this correlation becomes stronger with the incidence angle decreased. Our gradient of $V_{\phi}$ with $R$  for the thick disk is in good agreement with their low  orbital inclination stars, while our correlation between  $V_{\phi}$ and $|z|$ is consistent with high orbital inclination stars. It indicates that the dynamical heating model may not explain the shortage of the rotation velocity gradient with radial distance and vertical height.
 In addition, Figure \ref{figure14} shows that the absence of the peak at high e $\sim$ 0.8 in our orbital eccentricity distribution excludes dynamical heating model. We conclude that dynamical heating model could not have played a significant role but only involved in the process of formation and evolution of the thick disk.  In addition, the relative shortage of low eccentricity stars for our observation excludes gas-rich merger model, and our eccentricities distribution  is more consistent with accretion model but it also shows more stars at $0.3 < e < 0.6$ than accretion model prediction.  There could be several reasons for the accretion model not fitting the data. First, the thick disk may be formed by combined processes, 
 and other formation mechanisms  such as the radial migration or 
 heating scenario model also contribute to the distribution of orbital eccentricities.   
 Secondly, due to selective criterion,  the sample isn't complete enough to derive the distribution of orbit eccentricities for the thick disk stars.  
 Finally, it is possible that the simulation can't duplicate the formation of the Milky Way Galaxy exactly.
\par 
In summary, comparisons of our observed results with the predictions of the radial migration, gas-rich merger, accretion, and heating model suggest that radial migration may have influenced the structure and chemical evolution of the thin disk, but could not have played a  significant role in the formation  of the thick disk. The formation of the thick disk could be affected by more than one processes. The accretion model could play an indispensable role in the formation of the thick disk, and other formation mechanisms,  such as the radial migration or heating scenario model could also contribute to the formation of the thick disk.
\section{Summary and Conclusions}

\par Based on a sample of 307,246 giant stars from the LAMOST  spectroscopic survey and Gaia DR2 survey, 
which located at $ 4\lesssim R \lesssim 15$ kpc, extend up to 6 kpc in height from the Galactic plane, we investigate kinematics and metallicity distribution of the Galactic disk. First of all,  the sample stars are divided into the thin disk and thick disk components according to the chemical abundances ([Fe/H] and [$\alpha$/Fe]) and kinematics.  In total, we obtain 179,092 thin disk stars and 29,966 thick disk stars, and the metallicity distribution of the thin disk  can be described by Gaussian model with peaks at [Fe/H] $\sim -0.21$ and standard deviation $\sigma_{\rm [Fe/H]} \sim 0.20$.  The metallicity distribution of the thick disk has an extended metallicity tail, but it can be described by Gaussian model with peaks at [Fe/H] $\sim -0.52$, $\sigma_{\rm [Fe/H]} \sim 0.23$ in the range of [Fe/H] $>$ -1.2 dex.  
Our data has advantages of larger sample and wide spatial range for investigating chemistic and kinematics properties of the thin disk and thick disk.

\par For the thin disk, we summary the results as follows: \\
(1) The thin disk has a negative metallicity gradient with radial distance, but it becomes flat with increasing vertical height. While the inner disk ($R < 8$ kpc) of the thin disk has a positive metallicity gradient and the outer disk ($R > 8$ kpc) has a negative metallicity gradient with radial distance.\\
(2) The thin disk has a  negative metallicity gradient with vertical height, but it shows invariable in the inner region and then becomes flat with increasing radial distance.  Also, the thin disk has a positive [$\alpha$/Fe] gradient with vertical height.\\
(3) The thin-disk stars have low orbital eccentricities  (peak at $\sim0.12$). The orbital eccentricities exist a slight uptrend with increasing metallicity, and have no obvious relationship with vertical height.  The thin disk has a flat rotational velocity gradient with radial distance and vertical height, and it has a negative rotational velocity gradients with metallicity. \\

\par For the thick disk, there exists more controversies on chemical and kinematic properties.  
Our results are summarized as follows: \\
(1) The thick disk has no metallicity gradient with radial distance, while the inner disk ($R < 8$ kpc) has a positive metallicity gradient and the outer disk ($R > 8$ kpc) has a negative metallicity gradient.\\
(2) The thick disk has a negative metallicity gradient with vertical height, but it is slightly flat on average with increasing radial distance.
The thick disk has a positive [$\alpha$/Fe] gradient with vertical height.\\
(3) The orbital eccentricities of the thick disk  are higher, and its peaks at $\sim 0.42$, with wide widths, and extend up to $e \sim 0.8$. And, it exists a downtrend with increasing metallicity, and have no obvious relationship with vertical height.
The thick disk has a flat rotational velocity gradient with radial distance and vertical height,  but it has a positive rotational velocity gradient with metallicity.

Our results are in agreement with most previous studies including metallicity, [$\alpha$/Fe] and rotational velocity. More properties are also found for the thin disk and thick disk such as two different metallicity radial gradients for the inner and outer disk.  

\par

According to above derived chemical and kinematic properties in this study, we conclude that  radial migration could have influenced the structure and chemical evolution of the thin disk. But for the formation of the thick disk, it could be affected by more than one processes: the accretion model cloud play an indispensable role, and other formation mechanisms,  such as the radial migration or heating scenario model could also contribute to the formation of the thick disk.

\section{Acknowledgements}

\par We thank especially the referee for insightful comments and suggestions, which have improved the paper significantly. This work was supported by joint funding for Astronomy by the National Natural Science Foundation of China and the Chinese Academy of Science, under Grants U1231113.  This work was also by supported by the Special funds of cooperation between the Institute and the University of the Chinese Academy of Sciences, and China Scholarship Council (CSC).  In addition, this work was supported by the National Natural Foundation of China (NSFC No.11625313 and No.11573035).
The Guoshoujing Telescope (the Large Sky Area Multi-Object Fiber Spectroscopic Telescope, LAMOST) is a National Major Scientific Project built by the Chinese Academy of Sciences. Funding for the project has been provided by the National Development and Reform Commission. LAMOST is operated and managed by the National Astronomical Observatories, Chinese Academy of Sciences. This work has made use of data from the European Space Agency (ESA) mission Gaia (http://www.cosmos.esa.int/gaia), processed by the Gaia Data Processing and Analysis Consortium (DPAC, http://www.cosmos.esa.int/web/gaia/dpac/consortium). Funding for DPAC has been provided by national institutions, in particular the institutions participating in the Gaia Multilateral Agreement.

\end{document}